\documentclass[smus]{snow2e}
\usepackage{epsf}

\begin{document}

\title{Discovering New Interactions at Colliders}

\author{Kingman Cheung$^1$ and Robert M. Harris$^2$\\ 
$^1${\it University of Texas, Austin, TX \ \ 78712}\\
$^2${\it Fermilab, Batavia, IL \ \ 60510}}

\maketitle

\thispagestyle{empty}\pagestyle{empty}

\begin{abstract} 
%
We summarize results of the 1996 Snowmass workshop on future prospects for 
discovering dynamical electroweak symmetry breaking, compositeness, and 
anomalous couplings of quarks at colliders.
We present the mass reach of the Tevatron to a color singlet or octet 
technirho, and to a topgluon or topcolor Z$^{\prime}$ from topcolor assisted
technicolor. We explore the sensitivity of the Tevatron, LHC, NLC, and VLHC
to contact interactions and excited fermions. Finally we investigate the 
possibility of seeing anomalous couplings of quarks at the Tevatron and LHC.
\end{abstract}

\section{Dynamical Electroweak Symmetry Breaking}
The mechanism of electroweak symmetry breaking is unknown. The possibility
exists that electroweak symmetry is not broken by a fundamental higgs boson,
but instead is broken through the dynamics of a new interaction.  We explore
the discovery potential of future accelerators, and luminosity upgrades to 
the Tevatron, for two models of dynamical electroweak symmetry breaking: 
one-family technicolor and topcolor assisted technicolor.

\vspace*{-0.1in}
\subsection{One-Family Technicolor}

Eichten and Lane~\cite{ref_lane}  have presented a one-family technicolor 
model with color triplet
techniquarks and color singlet technileptons.  The techniquarks will bind to
form color singlet technirhos, $\rho_{T1}^{\pm}$ and $\rho_{T1}^0$, with mass
roughly in the range 200 to 400 GeV.  Color singlet technirhos 
are produced in hadron collisions through quark-antiquark annihilation.
The expected decay modes are 
$\rho_{T1}^\pm\rightarrow W^\pm Z$, $W^\pm\pi_T^0$, $Z\pi_T^\pm$, $\pi_T^\pm \pi_T^0$, 
and $\rho_{T1}^0 \rightarrow W^\pm W^\mp$, $W^\pm \pi_T^\mp$, $\pi_T^\pm \pi_T^\mp$. Here
the technipions, $\pi_T$, decay predominantly to heavy flavors: 
$\pi_T^0 \rightarrow b \bar{b}$, and $\pi_T^\pm \rightarrow c\bar{b}$, 
$t\bar{b}$.  Techniquarks will also bind to form color octet technirhos,
$\rho_{T8}^0$, with mass roughly in the range 200 to 600 GeV.  
Color octet technirhos are produced and decay via strong interactions.
If the mass
of the colored technipions is greater than half the mass of the technirho,
then the color octet technirho will decay predominantly to dijets:
$\rho_{T8} \rightarrow gg$.  If colored technipions are light
the color octet technirho decays to pairs of either color triplet 
technipions (leptoquarks) or color octet technipions.

\subsubsection{ $\rho_{T1} \rightarrow W$ + dijet at the Tevatron}

\begin{figure}[ht]
\epsfysize=3.0in
\hspace*{0.2in}
\epsffile{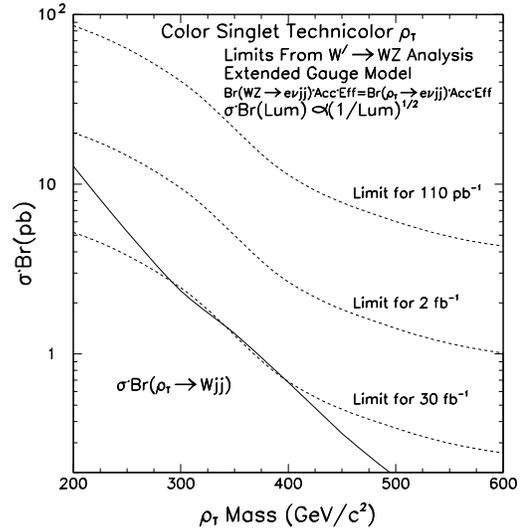}
\caption{95$\%$ CL upper limit of $\sigma \cdot Br(\rho_{T1} \rightarrow Wjj)$
vs.
M$_{\rho_T}$. 
The solid line is the theoretically expected $\sigma
\cdot Br$ and assumes $\rho_{T1}
\rightarrow WX \rightarrow Wjj = 100\%$. 
The dashed lines show predicted limits for 
110pb$^{-1}$, 2fb$^{-1}$ and 30fb$^{-1}$ respectively. Note that we have assumed
that the limits simply scale as the inverse of the square root of the
luminosity}
\label{fig_toback}
\end{figure}

The search for $\rho_{T1} \rightarrow W X$, where $X$ can be a $W$,$Z$, or 
$\pi_T$, is sufficiently similar to the search for a massive $W^{\prime}$
decaying to $WZ$, that Toback~\cite{ref_toback} has extrapolated the
$W^{\prime}$ search to higher luminosities as an estimate of our sensitivity to
color singlet technirhos at the Tevatron.  He considered the decay chain
$\rho_T \rightarrow W X \rightarrow e \nu +$ dijets, and required both the 
electron and neutrino to have more than 30 GeV of transverse energy, $E_T$.
He required at least two jets in the event, one with $E_T>50$ GeV, and the
other with $E_T>20$ GeV. The higher $E_T$ cut on the two jets was optimized 
for a high mass $W^{\prime}$ search ($M>500$ GeV) and should be reduced 
for a lower mass technirho search.  The resulting W+dijet mass distribution 
from 110 pb$^{-1}$ of CDF data was in good agreement with standard model 
predictions, and was used to determine the 95\% CL upper limit on the 
$\rho_{T1}$ cross section, shown in Fig.~\ref{fig_toback}. Here he assumed
that the acceptance for a technirho was roughly the same as for a $W^{\prime}$.
The extrapolation to higher luminosities shows that TeV33 (30 fb$^{-1}$) should 
be able to exclude at 95\% CL a color singlet technirho decaying to W plus 
dijets for technirho masses up to roughly 400 GeV.  This covers the expected
range in the one-family technicolor model.

\vspace*{-0.3in}

\vspace*{-0.2in}
\subsubsection{$\rho_{T1} \rightarrow W + b\bar{b}$ at the Tevatron and LHC}
\begin{figure}[tbh]
\vspace*{2.2in}
\includegraphics{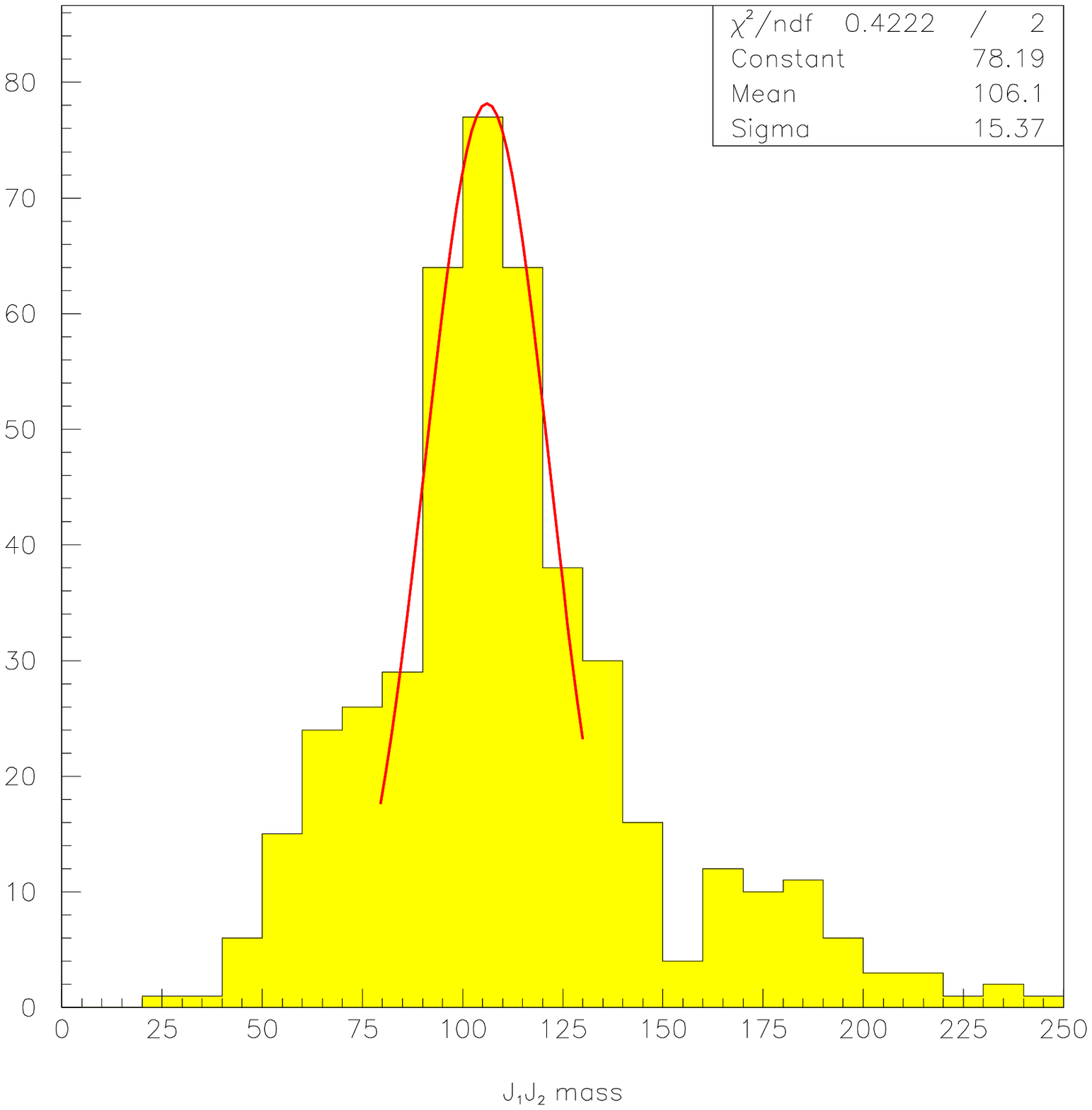}
\includegraphics{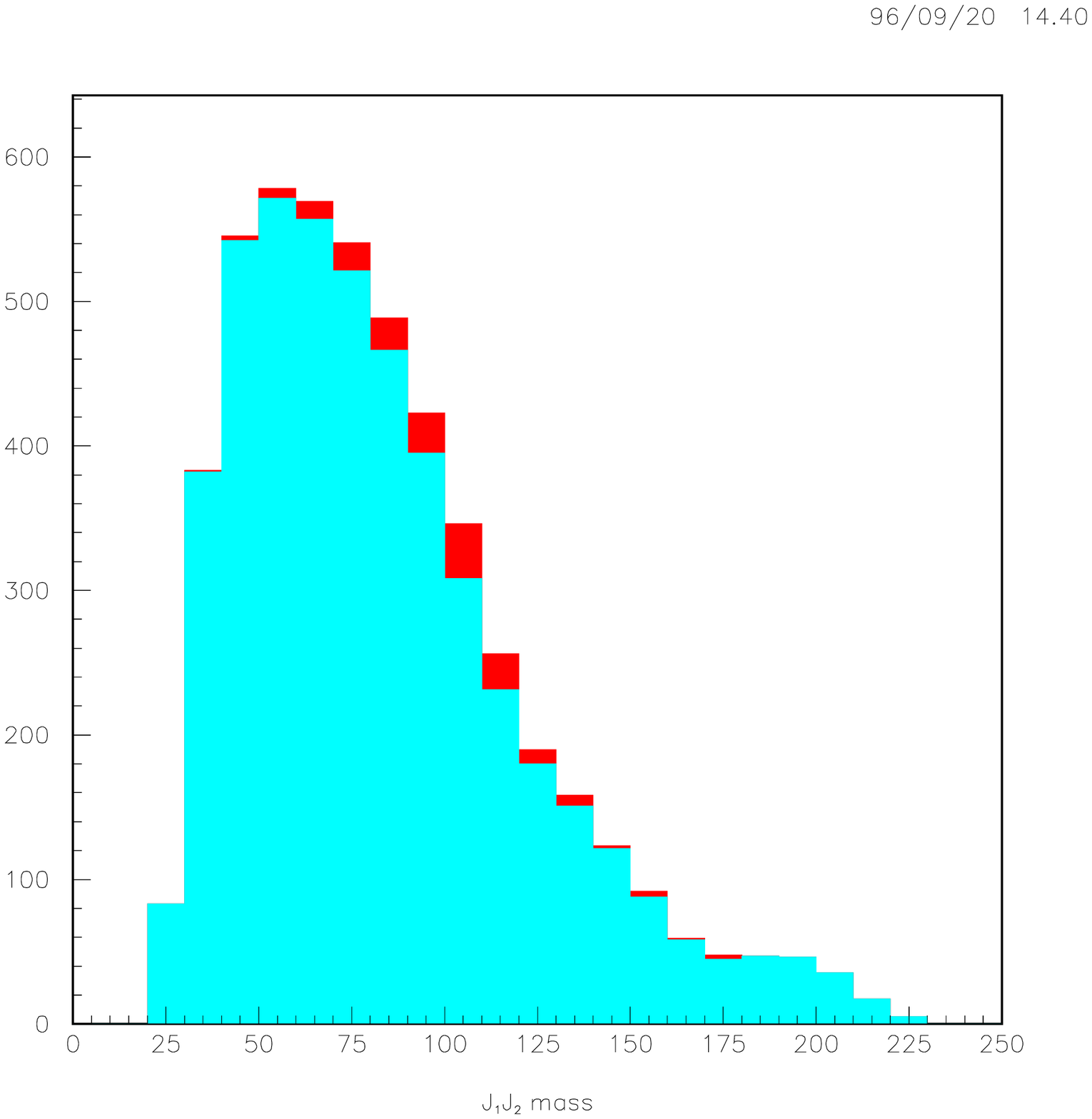}
\includegraphics{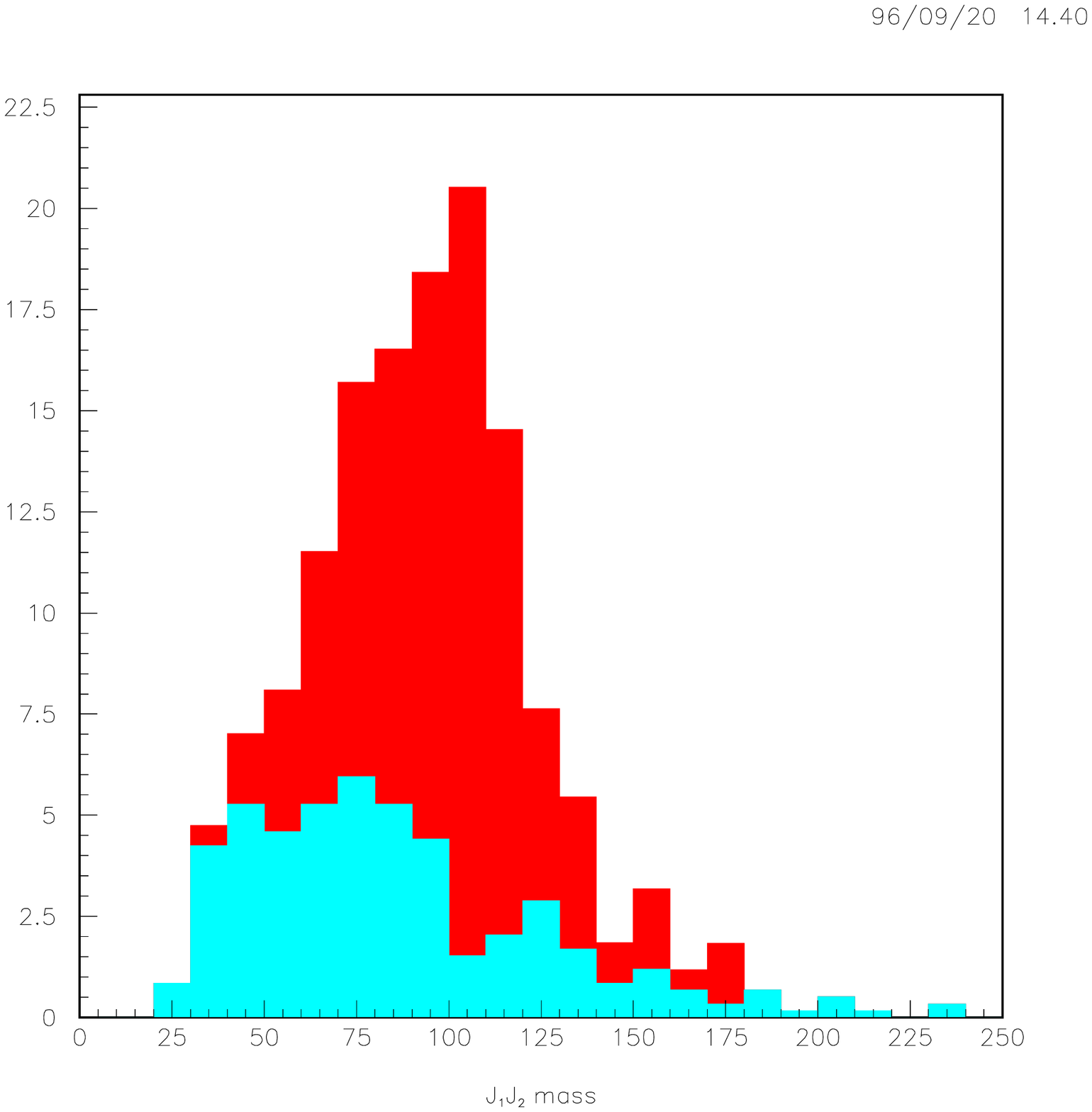}
\includegraphics{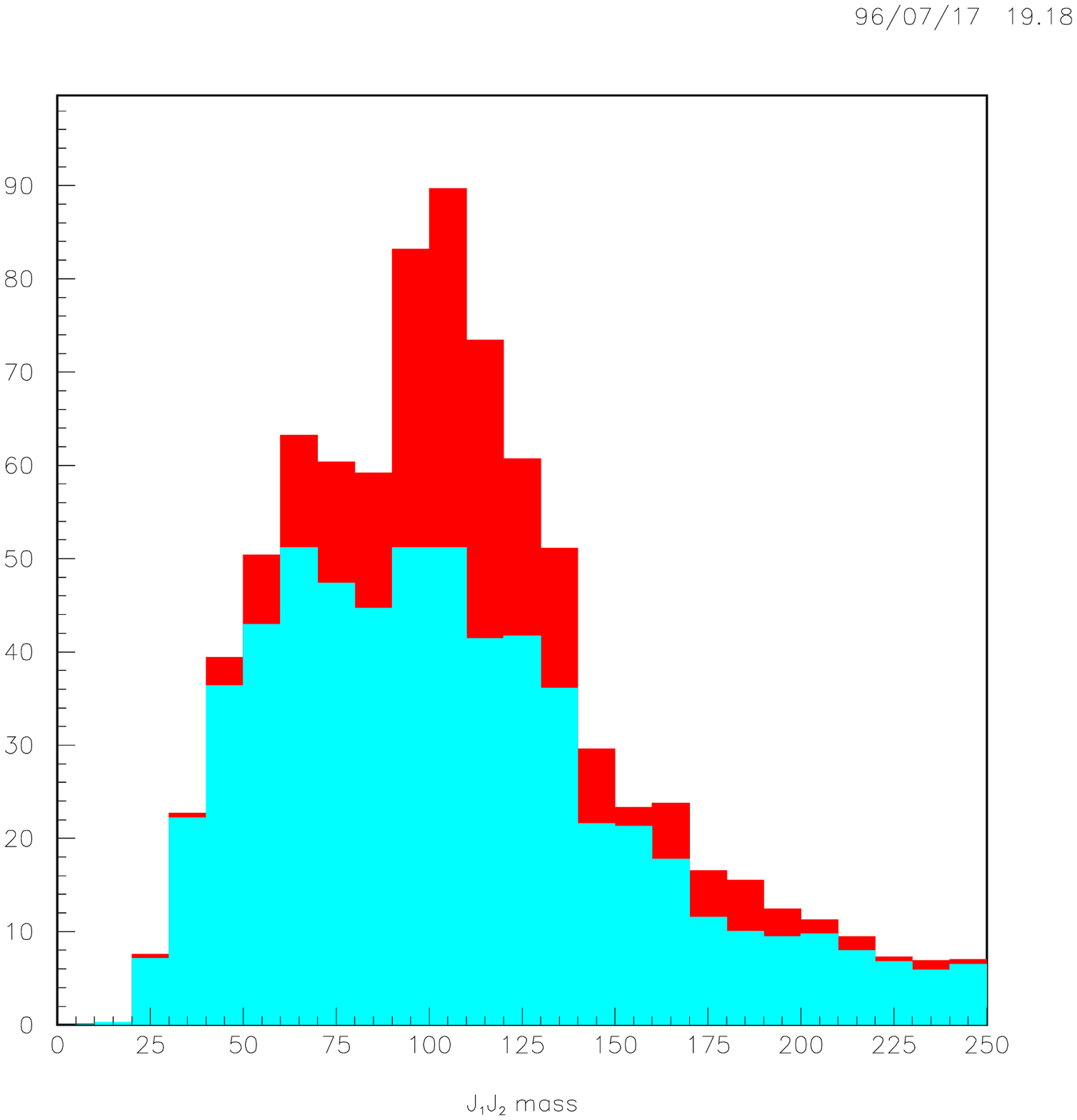}
\vspace*{1.3in}
\caption{$\rho_{T1} \rightarrow W + \pi_T \rightarrow (l\nu)(b\bar{b})$ 
search. (upper left) Leading dijet invariant mass distribution for signal at 
the LHC. (upper right) Same for signal (dark) and background (light) at the 
Tevatron before b-tagging. Vertical scale is events/10 GeV/2 fb$^{-1}$.  
(lower left) Same at the Tevatron after b-tagging. (lower right) Same at the
LHC after b-tagging. Vertical scale is events/10 GeV/0.5 fb$^{-1}$. All 
horizontal scales are in GeV.}
\label{fig_womersley}
\end{figure}

Womersley~\cite{ref_womersley} has studied the process 
$q\bar{q}^{\prime} \rightarrow \rho_{T1} \rightarrow W\pi_T \rightarrow (l\nu)(b\bar{b})$,
including the effect of tagging events with a final state $b$ quark, for the 
particular 
case of $m_{\rho_T} = 210$ GeV and $m_{\pi_T}=115$ GeV. He generates signal and
background events using ISAJET, and uses a fast simulation of the CMS detector 
at the LHC. After all simulation, events are required to have a good W
candidate, formed from an isolated charged lepton with $E_T>25$ GeV and 
pseudorapidity $|\eta|<1.1$, a neutrino with $E_T>25$ GeV, and their combined
transverse mass in the range $50 < m_T < 100$ GeV.  Further, events were 
required to have two jets with $E_T>20$ GeV and $|\eta|<2.5$, and the 
probability of tagging at least one of the two b quarks was assumed to be
50\% with a mistag rate of 1\% for light quarks. Figure~\ref{fig_womersley}
show the reconstructed $\pi_T$ peak in the signal sample, and that prior
to b-tagging the signal is swamped by a large QCD W+dijet background. 
Figure~\ref{fig_womersley} also shows that after b-tagging the signal to 
background is significantly improved at both the Tevatron and the LHC. For 
this particular case of a light technirho the signal to background is better
at the Tevatron although the rate at the LHC is considerably higher. 
Clearly, b-tagging is critical, and makes possible the discovery of a 210 GeV 
color singlet technirho at the Tevatron in Run II (2 fb$^{-1}$).

\vspace*{-0.2in}
\subsubsection{$\rho_{T8} \rightarrow$ dijets at the Tevatron}
\begin{figure}[tbh]
\vspace*{-.1in}
\epsfysize=3.5in
\epsffile{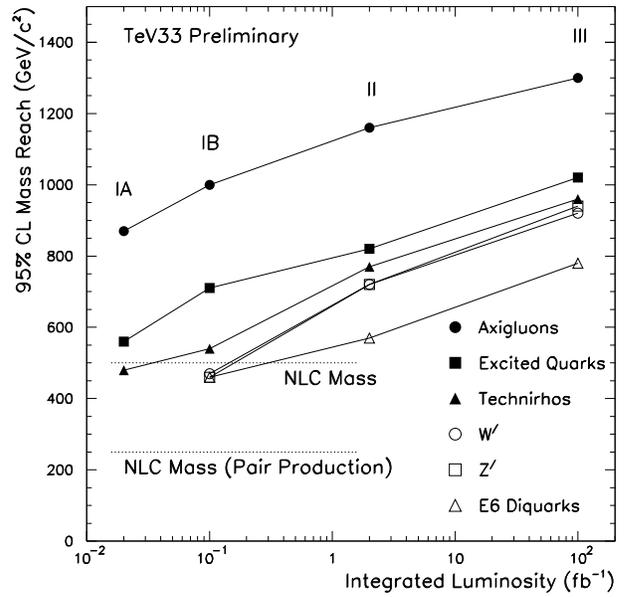}
\caption{The mass reach for new particles decaying to dijets 
vs. integrated luminosity at the Tevatron.
The mass reach of the NLC for direct production is also shown.}
\label{fig_dijet}
\end{figure}

Harris has determined the sensitivity at the Tevatron to dijet decays of color 
octet 
technirhos by extrapolating CDF searches~\cite{ref_dijet} to higher 
luminosities. Here 
there are significant QCD backgrounds, so the cross section limits scale 
inversely as the square root of the luminosity. In reference~\cite{ref_tev2000}
he compared the cross section limit to the theoretical prediction, to determine
the mass excluded at 95\% CL, shown in fig.~\ref{fig_dijet}. The mass reach
for color octet technirhos is $0.77$ TeV for Run II (2 fb$^{-1}$) and $0.90$ TeV
for TeV33 (30 fb$^{-1}$), which is more than the expected $\rho_{T8}$ mass in
the one-family technicolor model.

\vspace*{-0.3in}
\subsubsection{$gg\rightarrow Z_L Z_L, W_L W_L$ at LHC}
\begin{figure}[tbh]
\vspace*{1.7in}
\includegraphics{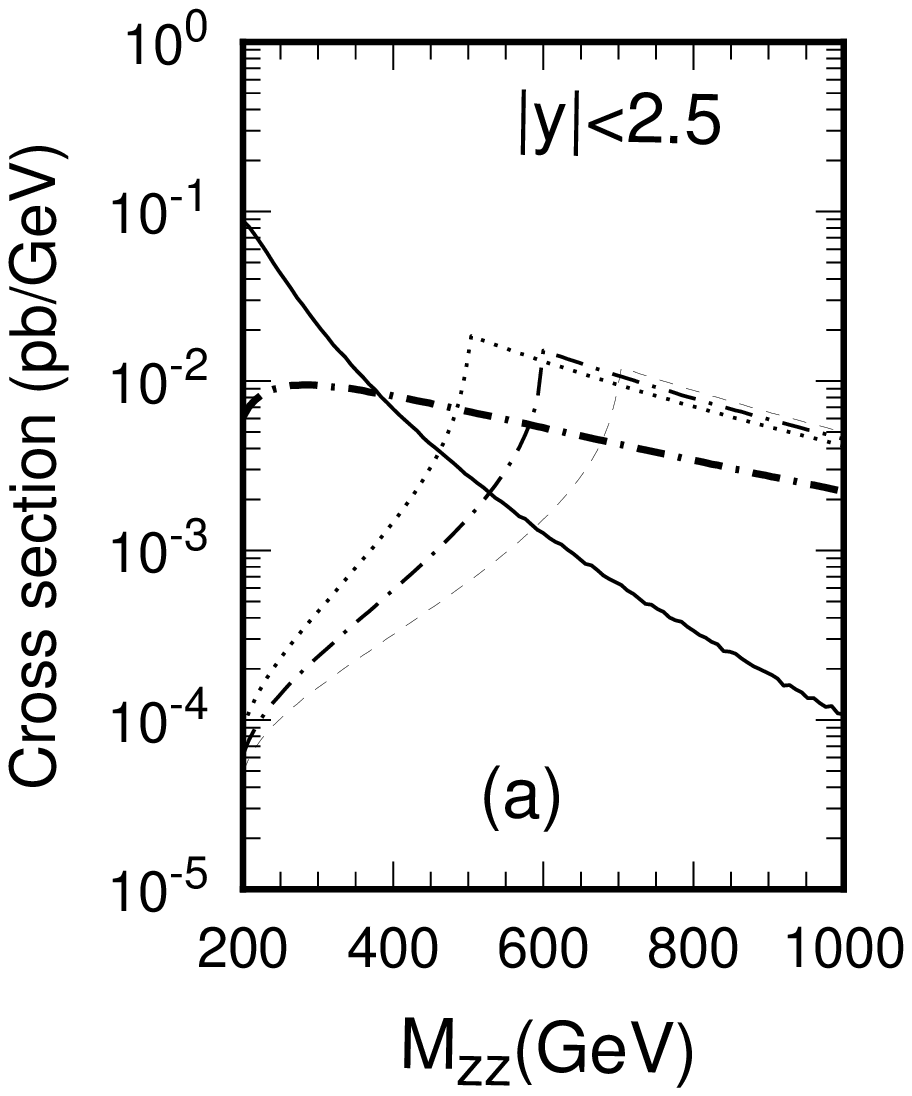}
\includegraphics{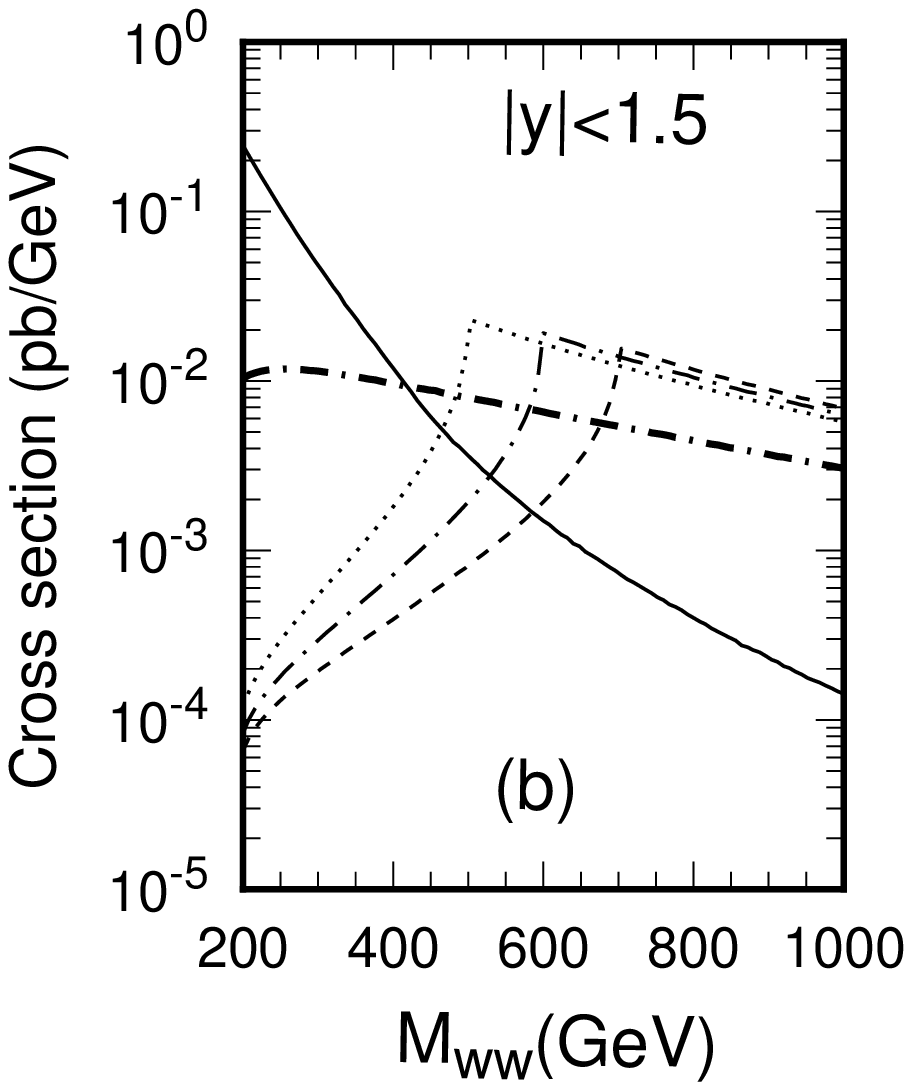}
\caption{The cross sections for {\bf a)} $Z_{L}$-pair and {\bf b)} $W_{L}$-pair production
via gluon fusion in proton collisions at $E_{c.m.}= 14$ TeV. The solid curves 
are for the $q \bar{q}$ initiated backgrounds, and dotted, dot-dashed, and dashed
curves are for technipion masses
 of $250 \,GeV$, $300 \, GeV$, and $350 \, GeV$ respectively. The thick
\noindent dot-dashed curves are for the chiral limit ($m_{\pi_T}=0$).}
\label{fig_taekoon}
\end{figure}

Lee~\cite{ref_taekoon} has studied the production of longitudinal weak gauge
boson pairs via gluon fusion in a one-family technicolor model~\cite{ref_lane} 
at the LHC. Fig.~\ref{fig_taekoon} shows that when the invariant 
mass is above the threshold for production of pairs of colored technipions, the 
$W_L W_L$ or $Z_L Z_L$ signal cross section is greater than the standard 
model background by over
an order of magnitude.  Assuming an integrated luminosity of 100 fb$^{-1}$,
the $Z_L Z_L$ signal, over a thousand events with four 
leptons in the final state (e and $\mu$), will be easily observable. 
If one-family technicolor exists, the LHC will see it in this channel.

\subsection{Topcolor Assisted Technicolor}
Eichten and Lane~\cite{ref_lane2} have recently discussed the phenomenology of 
the topcolor model of Hill and Parke~\cite{ref_topcolor}, and Burdman
has recently studied the scalar sector of the model~\cite{ref_burdman}. 
Topcolor assisted technicolor~\cite{ref_hill} is a model of dynamical
electroweak symmetry breaking in which the top quark is heavy because of a
new dynamics. Topcolor replaces the $SU(3)_C$ of QCD with $SU(3)_1$ for 
the third quark generation and $SU(3)_2$ for the first two generations. 
The additional SU(3) symmetry produces a $<t\bar{t}>$ condensate 
which makes the top quark heavy, and gives rise to a color octet gauge
boson, the topgluon B. The topgluon is expected to be wide 
($\Gamma/M \approx 0.3 - 0.7$) and 
massive ($M \sim 0.5 - 2$ TeV). In hadron collisions it is produced through 
a small coupling to the first two generations, and then decays via a much larger
coupling to the third generation: $q\bar{q} \rightarrow B \rightarrow 
b\bar{b}, t\bar{t}$. 

Similarly, topcolor also replaces 
$U(1)_Y$ of the standard model with $U(1)_{Y1}$ for the third generation and 
$U(1)_{Y2}$ for the first two 
generations. The additional $U(1)$ keeps the bottom quark light, and gives 
rise to a massive color singlet gauge boson, the topcolor $Z^\prime$.  The
topcolor $Z^{\prime}$ may be narrow ($\Gamma/M \sim 0.01 - 0.1$) and it couples 
predominantly to $t\bar{t}$.

\vspace*{-0.1in}
\subsubsection{Topgluons decaying to $b\bar{b}$ at the Tevatron}

\begin{figure}[hb!]
\vspace*{-0.31in}
\epsfysize=3.4in
\epsffile{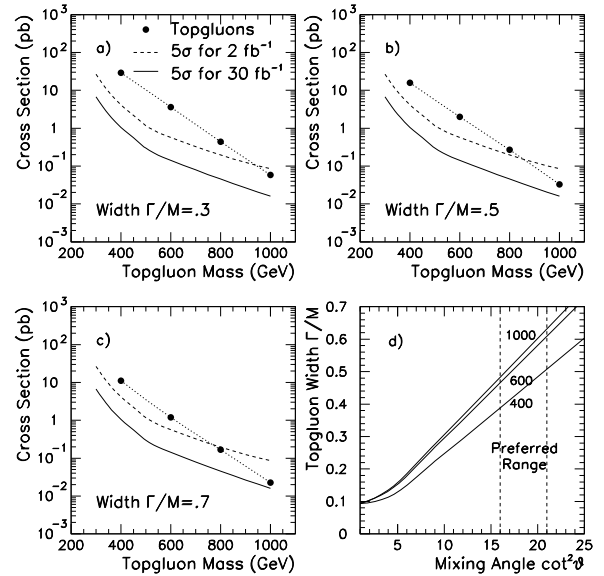}
\vspace*{-0.2in}
\caption[]{ The mass reach for $b\bar{b}$ decays of topgluons of width a)
0.3 M, b) 0.5 M, and c) 0.7 M.  The cross section for
topgluons (points) is compared to the 5$\sigma$ discovery reach of the Tevatron
with a luminosity of 2 fb$^{-1}$ (dashed) and 30 fb$^{-1}$ (solid). 
d) Topgluon width as a function of mixing angle between $SU(3)_1$ and 
$SU(3)_2$ for 3 topgluon masses (curves). The vertical dashed lines are
the theoretically preferred range of mixing angle~\cite{ref_topg_range}.}
\label{fig_topg_bbbar}
\end{figure}

Harris~\cite{ref_topg_bbbar} has used a full simulation of topgluon 
production and decay to $b\bar{b}$,
and an extrapolation of the b-tagged dijet mass data~\cite{ref_pbarp}, to
estimate the topgluon discovery mass reach in a $b\bar{b}$ resonance search. 
Fig.~\ref{fig_topg_bbbar} displays the results for three
different widths of the topgluon. The topgluon discovery mass reach, 
$0.77-0.95$ TeV for Run II and $1.0-1.2$ TeV for TeV33, covers a significant 
part of the expected mass range ($\sim0.5 - 2$ TeV).  

\begin{figure}[th!]
\vspace*{-0.1in}
\epsfysize=3.4in
\epsffile{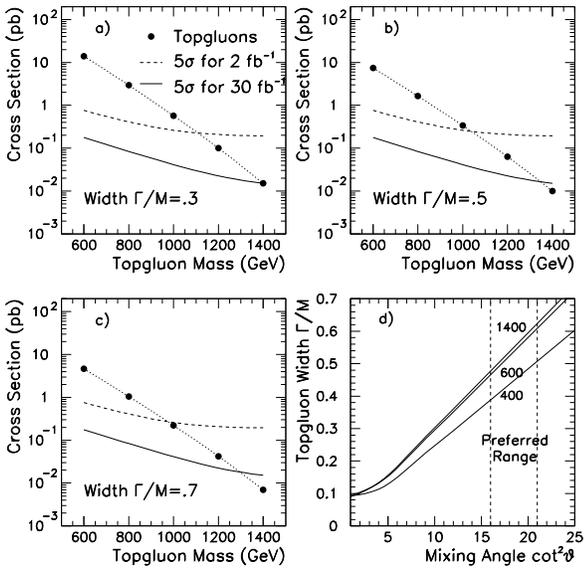}
\caption[]{ Same as Fig.~\ref{fig_topg_bbbar} for $t\bar{t}$ decays of
topgluons.}
\label{fig_topg_ttbar}
\end{figure}

\subsubsection{Topgluons decaying to $t\bar{t}$ at the Tevatron}

\begin{figure}[tbh]
\vspace*{-.15in}
\epsfysize=3.5in
\epsffile{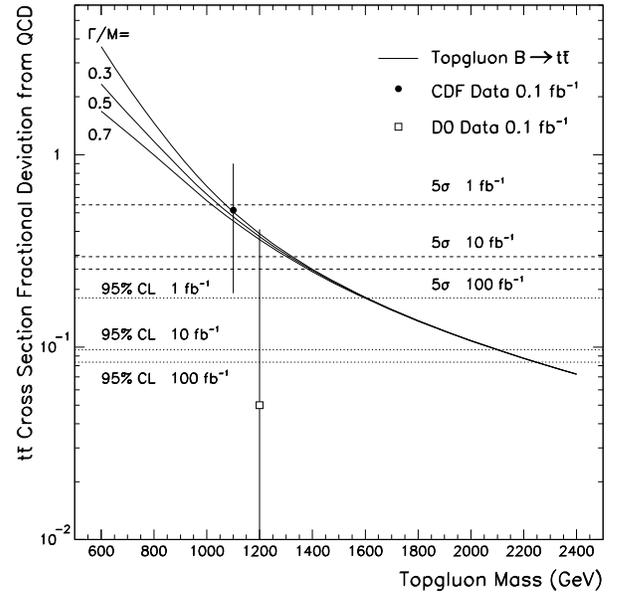}
\vspace*{-.15in}
\caption[]{ The fractional difference between the $t\bar{t}$ cross section
and the QCD prediction is shown for topgluons (solid
curves), CDF data (solid circle), and D0 data (open box).
The projected $5\sigma$ uncertainty (dashed lines) and 95\% CL (dotted lines)
on the measured $t\bar{t}$ cross section can be compared with the topgluon
prediction to determine the discovery reach and exclusion reach of the
Tevatron at the luminosities of 1, 10 and 100 fb$^{-1}$.}
\label{fig_ttbar_xsec}
\end{figure}

Harris~\cite{ref_topg_ttbar} has used a parton level prediction for 
$t\bar{t}$ production from QCD
and topgluons, together with the projected experimental efficiency for 
reconstructing $t\bar{t}$,
to estimate the topgluon discovery mass reach in a $t\bar{t}$ resonance search. 
Fig.~\ref{fig_topg_ttbar} displays the results for three
different widths of the topgluon. 
The topgluon discovery mass reach,
$1.0-1.1$ TeV for Run II and $1.3-1.4$ TeV for TeV33, covers a significant part 
of the expected mass range ($\sim0.5 - 2$ TeV).  
The mass reach estimated using the total $t\bar{t}$ cross section, shown
in Fig.~\ref{fig_ttbar_xsec}, is similar
to that for the resonance search, providing an important check.
This mass reach is better than in the $b\bar{b}$ channel, discussed in
the previous section, because backgrounds in the $t\bar{t}$ channel are smaller.
If topgluons exist, there is a good chance we will find
them at the Tevatron.

\begin{figure}[bh!]
\vspace*{-.15in}
\epsfysize=3.5in
\epsffile{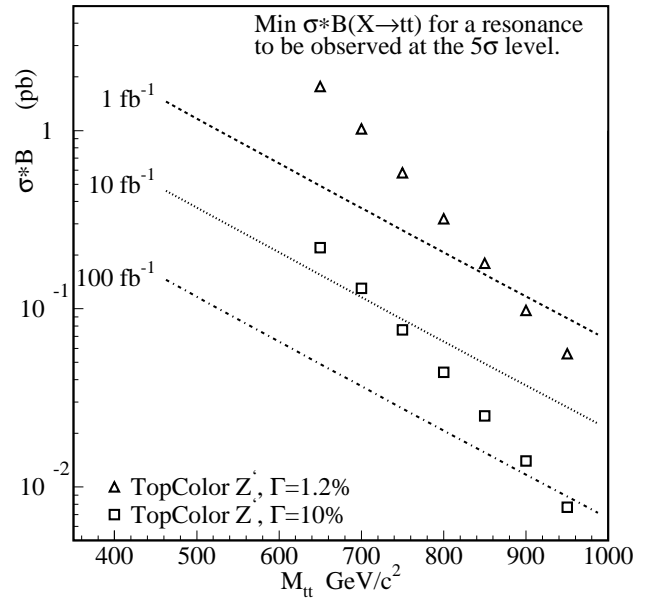}
\vspace*{-.15in}
\caption[]{ $\sigma . B(X\rightarrow t\bar{t})$ vs. $t\bar{t}$ mass. 
The minimum cross section to observe a $5\sigma$ excess of
events in a sample with 1, 10 and 100 fb$^{-1}$ (lines) is compared to 
the expected cross section for a topcolor $Z^{\prime}$ with width 
$\Gamma/M=.012$ (triangles) and $\Gamma/M=.1$ (squares).}
\label{fig_zprime}
\end{figure}

\subsubsection{Topcolor $Z^{\prime}$ decaying to $t\bar{t}$ at the Tevatron}

Tollefson~\cite{ref_tollefson} has considered the decay chain topcolor 
$Z^{\prime} \rightarrow t\bar{t} \rightarrow (Wb)(W\bar{b})\rightarrow l\nu b \bar{b}jj$.
She uses the PYTHIA Monte Carlo~\cite{ref_pythia} and a CDF detector simulation for both the
signal and background. As in the CDF top quark mass 
analysis~\cite{ref_cdf_top}, she requires a central charged lepton with 
$E_T>20$ GeV, a neutrino with $E_T>20$ GeV, 3 jets with $E_T>15$ GeV and 
$|\eta|<2$, one jet with $E_T>8$ GeV and $|\eta|<2.4$ and requires that at least
two of the four jets be tagged as a $b$ quark.  She reconstructs the $t\bar{t}$ 
mass with the following mass constraints: the charged lepton and 
neutrino reconstruct to the mass of a W, the two jets reconstruct to the mass
of a W, and the mass of each reconstructed top quark be 175 GeV. This
results in $t\bar{t}$ mass resolution of 6\% and an acceptance of 6.5\% for
the signal.  From a binned maximum likelihood fit of the simulated $t\bar{t}$ 
mass distribution, she determines what resonance cross section would produce
a 5$\sigma$ signal, and compares that to the expected topcolor $Z^{\prime}$
cross section in Fig.~\ref{fig_zprime}.  The resulting mass reach for
a narrow topcolor $Z^{\prime}$ at the Tevatron is $0.9$ TeV for Run II 
(2 fb$^{-1}$) and $1.1$ TeV for TeV33 (30 fb$^{-1}$).



\section{Composite Fermions}

The repetition of the three generations of quarks and leptons strongly
suggests that they are composite structures made up of more fundamental
fermions, which are often called ``preons'' in the literature.  There have
been a lot of theoretical efforts to construct realistic models for
composite fermions, but no obviously correct or compelling model exists.
Nor do we know the energy scale $\Lambda$ which characterizes the 
interactions of preons.

\subsection{Contact Interactions}

%
Deviations from the 
Standard Model (SM) in low energy phenomena can be systematically studied 
using the effective Lagrangian approach.
In this approach, an effective Lagrangian, which obeys the low energy SM
symmetries, is constructed out of the SM fields.
The leading terms are simply given by the SM, 
while the higher order terms consist of 
higher-dimension operators and are 
suppressed by powers of the scale $\Lambda$ of 
the new physics.  

The existence of 
quark and lepton substructure will be signaled by the appearance of the 
four-fermion contact interactions
at energies below $\Lambda$ \cite{ekp}.  Eichten and Lane have reviewed these
contact interactions~\cite{ref_lane2}.
They arise from the exchanges of preon bound states, and they must be
$SU(3) \otimes SU(2) \otimes U(1)$ invariant because they are generated by
forces operating at or above the electroweak scale.
The lowest order four-fermion contact interactions are of dim-6, which means
that they are suppressed by $1/\Lambda^2$.
The general Lagrangian of four-fermion contact interactions, up to dimension 6, 
can be written as
\begin{equation}
{\cal L}  \sim \frac{g^2\eta}{2\Lambda^2} \biggr( \bar q \gamma^\mu q + 
 {\cal F}_\ell \bar \ell \gamma^\mu \ell \biggr )_{L/R}
\; \biggr( \bar q \gamma_\mu q + 
 {\cal F}_\ell \bar \ell \gamma_\mu \ell \biggr )_{L/R} \;
\end{equation}
where we have suppressed the generation and color indices, $\eta=\pm 1$, and
${\cal F}_\ell$ is inserted to allow for different quark and lepton couplings
but is expected to be ${\cal O}(1)$.   It is conventional to define $g^2=4\pi$, 
so that the interaction is defined to be strong when $\hat{s}$ approaches 
$\Lambda$.
These contact interactions can affect jet production, the Drell-Yan
process, lepton scattering, etc.
Since compared to the SM the contact interaction amplitudes are of order
$\hat s/\alpha_S \Lambda^2$ or $\hat s/\alpha_{\rm em} \Lambda^2$, the effects
of the contact interactions will be most important in the phase space region
with large $\hat s$.   Therefore, the four-fermion contact interactions
are often searched for at the high $E_T$ region in jet and lepton-pair
production.  
So far, the contact interaction used most to parameterize the substructure
scale $\Lambda$, is the product of two left-handed electroweak isoscalar
quark and lepton currents.  

\subsubsection{$l\bar{l}\rightarrow q\bar{q}$ and $l\bar{l} \rightarrow 
l^{\prime}\bar{l}^\prime$ Contact}

\begin{table}[th!]
\vspace*{-.1in}
\begin{center}
\caption{95\% CL lower bounds on $\Lambda$ at lepton colliders, 
as a function of center of mass energy and integrated luminosity, shown for 
each possible helicity of the interaction.}
\label{table-cgh}
\begin{tabular}{|@{\extracolsep{-0.06in}}cc|l|cccc|}
\hline
\hline
\multicolumn{7}{|c|}{$e^+e^-$ Colliders}\\ \hline
$\sqrt{s}$  & $L$  & & \multicolumn{4}{c|}{$\Lambda$ in TeV}  \\ 
\multicolumn{1}{|c}{TeV} & \multicolumn{1}{c|}{fb$^{-1}$}
& \multicolumn{1}{c|}{Process}  & $LL$ & $LR$ & $RL$ & $RR$ \\ 
\hline
0.5 & 50 &   $e^- e^+ \to \mu^+ \mu^-$  & 19 & 16 & 16 & 18 \\
    &  &   $e^- e^+ \to b \bar{b}$      & 24 & 18 & 5.5 &16 \\
   &  &   $e^- e^+ \to c \bar{c}$   & 20  & 4.2 &  5.6 & 17 \\ 
\hline
1.0 &200 & $e^- e^+ \to \mu^+ \mu^-$ & 37 & 33 &  33 &  36 \\
       & & $e^- e^+ \to b \bar{b}$ &   48 & 36 &  11 &  33 \\
       & & $e^- e^+ \to c \bar{c}$ & 5.1 &  8.3 & 11 &  5.9 \\
\hline 
1.5&200 & $e^- e^+ \to \mu^+ \mu^-$ &  45 & 40 & 40 &  44 \\
        & & $e^- e^+ \to b \bar{b}$ &  59 & 44 & 17 &  41 \\
        & & $e^- e^+ \to c \bar{c}$ &   7.7 &  12 & 16 &  8.9 \\
\hline
5.0&1000 & $e^- e^+ \to \mu^+ \mu^-$ & 120 &   110 &   110 &   120 \\
         & & $e^- e^+ \to b \bar{b}$ & 160 &   120 &   54 &   110 \\
         & & $e^- e^+ \to c \bar{c}$ & 26 &    40 &   51 &   29 \\
\hline \hline
\multicolumn{7}{|c|}{$\mu^+\mu^-$ Colliders}\\ \hline
0.5 & 0.7 & $\mu^- \mu^+ \to \tau^+ \tau^-$ & 6.3 & 5.7 & 5.7 & 6.1 \\
     & & $\mu^- \mu^+ \to b \bar{b}$ &  8.0 & 6.3 & 4.3 & 6.1 \\
     & & $\mu^- \mu^+ \to c \bar{c}$ &  6.9 & 3.5 & 4.0 &  2.7 \\
\hline
0.5 & 50 & $\mu^- \mu^+ \to \tau^+ \tau^-$ & 19 & 16 & 16 &  18 \\
             & & $\mu^- \mu^+ \to b \bar{b}$ & 24 & 18 &  5.5 & 16 \\
             & & $\mu^- \mu^+ \to c \bar{c}$ & 20 & 4.2 & 5.6 & 17 \\
\hline
4.0 & 1000 & $\mu^- \mu^+ \to \tau^+ \tau^-$ & 110 & 99 &  99 & 110 \\
       & & $\mu^- \mu^+ \to b \bar{b}$ & 140 &   110 &   44 & 100 \\
       & & $\mu^- \mu^+ \to c \bar{c}$ & 20 &   33 &   42 &  24 \\
\hline \hline
\end{tabular}
\end{center}
\vspace*{-0.15in}
\end{table}

Cheung, Godfrey, and Hewett \cite{cgh} studied the $\ell\ell qq$ and
$\ell\ell\ell'\ell'$ contact interactions at future $e^+e^-$ and
$\mu^+\mu^-$ colliders, and derived limits on the compositeness mass
scale $\Lambda$ using the reactions
$\ell^+ \ell^- \to f\bar{f}$,
 where $f=\mu,\tau,b,c$ and $\ell=e,\mu$ ($\ell \not = f)$.
These reactions proceed via $s$-channel exchanges of 
$\gamma$, $Z$, and the $\ell\ell f \bar f$ contact interaction.
The polarized differential cross sections for $e^-_{L/R} e^+ \to f\bar f$
versus $\cos\theta$, where $\theta$ is the scattering angle in the CM frame, 
are given by
\begin{equation}
{{d\sigma_L} \over {d\cos\theta}} = {{\pi \alpha^2 C_f}\over{4s}}
\left\{ |C_{LL}|^2(1+\cos\theta)^2 +|C_{LR}|^2 (1-\cos\theta)^2 \right\}
\end{equation}
where
\begin{equation}
C_{LL} = -Q_f +{{C_L^e C_L^f}\over {c_w^2 s_w^2}} 
{s\over{s-M_Z^2+i\Gamma_Z M_Z}} 
+{{s\eta_{LL}}\over{2\alpha\Lambda^2}}
\end{equation}
\begin{equation}
C_{LR} = -Q_f +{{C_L^e C_R^f}\over {c_w^2 s_w^2}} 
{s\over{s-M_Z^2+i\Gamma_Z M_Z}} 
+{{s\eta_{LR}}\over{2\alpha\Lambda^2}}
\end{equation}
and $C_L^f=T_{3f} - Q_f s_w^2$, $C_R^f= - Q_f s_w^2$, $C_f=3(1)$ for $f$
being a quark (lepton), $s_w$ and $c_w$ are, respectively, the sine and
cosine of the weak mixing angle.
The expressions for $d\sigma_R/d\cos\theta$, $C_{RR}$, and 
$C_{RL}$ can be obtained by interchanging $L \leftrightarrow R$.
The unpolarized differential cross section is simply given by the 
average of $d\sigma_L/d\cos\theta$ and $d\sigma_R/d\cos\theta$.
Other observables, e.g., $A_{FB}, A_{LR}$, can be obtained from these $\cos
\theta$ distributions.

To obtain the  sensitivity to the compositeness scale
they  assume that the SM is correct and perform a $\chi^2$ analysis of the
$\cos\theta$  distribution for the theory with a finite $\Lambda$.
An acceptance cut $|\cos\theta|<0.9$ was imposed and the whole $\cos\theta$
distribution is divided into 10 equal bins.
The efficiencies in detecting the final state are $\epsilon= 60\%$ for 
$b$ quarks, 35\% for $c$ quarks, and 100\% for leptons.
The limits on $\Lambda$ at 95\% CL that can be obtained by various processes
at future $e^+e^-$ and $\mu^+\mu^-$ colliders are 
tabulated in Table~\ref{table-cgh}.  
Very substantial improvements in probing
the compositeness mass scale can be achieved.  A 0.5 TeV $e^+e^-$ collider
with a $50\; {\rm fb}^{-1}$ luminosity can probe up to around 20 TeV, which
is better than Run II of the Tevatron.  Up to about 40, 60, and
160 TeV can be probed at $\sqrt{s}=1.0, 1.5$, and 5.0 TeV $e^+e^-$ machines, 
respectively. A 4 TeV $\mu^+\mu^-$ collider, which is under intensive study, 
can probe up to about 140 TeV.  Slightly better results can be obtained by 
using polarized $e^-$ beams with the same luminosity \cite{cgh}.

\subsubsection{$q\bar{q}\rightarrow l\bar{l}$ Contact}
P. de Barbaro {\em et al}~\cite{ref_barbaro} have studied the effect of a 
left-handed contact interaction between quarks and leptons at the Tevatron.  
Using 110 pb$^{-1}$ of CDF data on dielectron production, they report 
preliminary limits of $\Lambda_{LL}^-(q\bar{q}\rightarrow e^+e^-) \geq 3.4$ TeV
and $\Lambda_{LL}^+(q\bar{q}\rightarrow e^+e^-) \geq 2.4$ TeV at 95\% CL.
They also report limits for the dimuon channel and the combined 
dielectron+dimuon channels; the latter is approximately $0.5$ TeV more 
stringent than with electrons alone.  Using a Monte Carlo procedure they
estimate the sensitivity of the Tevatron with higher luminosities.  

\begin{figure}[tbh]
\hspace*{0.25in}
\epsfysize=3in
\epsffile{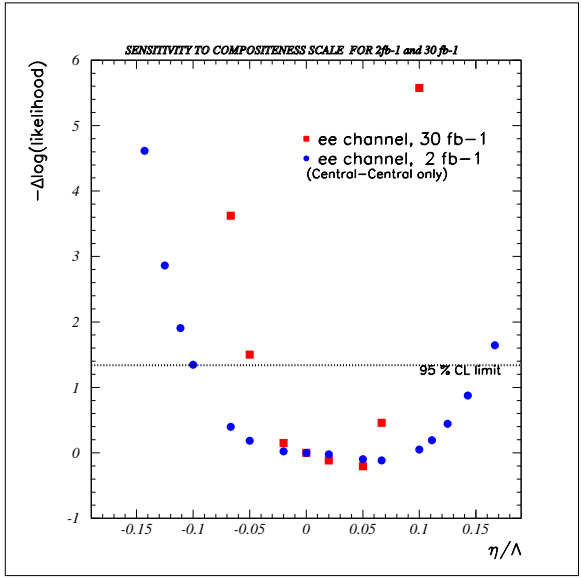}
\caption{
The change in the log likelihood function from the maximum plotted as a 
function of $\eta \times 1/\Lambda$ for $ee$ channel with 2 fb$^{-1}$ 
(circles) and 30 fb$^{-1}$ (squares).
The 95\% CL one sided limit occurs where the solid line at 1.34 intersects the 
points.}
\label{fig_likelihood}
\end{figure}

For standard model production of dielectrons they simulate one hundred
experiments with 2 fb$^{-1}$ and one hundred experiments with 30 fb$^{-1}$, 
each measuring the dielectron mass spectrum. For each
experiment they calculate a likelihood as a function of $\Lambda$ of that
experiment coming from the standard model plus a contact interaction of
strength $\Lambda$. To minimize fluctuations in the shape of the likelihood
function, they average the likelihood functions from the 100 experiments.
In Figure~\ref{fig_likelihood} they plot the log likelihood as a function of 
$\eta/\Lambda$, where $\eta$ is the sign of the contact interaction.  From
Fig.~\ref{fig_likelihood} a 
Tevatron experiment with 2 fb$^{-1}$ would exclude 
$\Lambda_{LL}^-(q\bar{q}\rightarrow e^+e^-) \leq 10$ TeV and 
$\Lambda_{LL}^+(q\bar{q}\rightarrow e^+e^-) \leq 6.5$ TeV,
and 30 fb$^{-1}$ would exclude 
$\Lambda_{LL}^-(q\bar{q}\rightarrow e^+e^-) \leq 20$ TeV and 
$\Lambda_{LL}^+(q\bar{q}\rightarrow e^+e^-) \leq 14$ TeV. The sensitivity is
always greater for $\eta=-1$, because this corresponds to constructive 
interference between the standard model and the contact interaction, and hence
a larger number of dielectrons.

\subsubsection{$qq \rightarrow qq$ Contact}

An excess of events with high jet $E_T$ in hadron collisions is a 
well known signature for a $qq \rightarrow qq$ contact interaction.  However, 
significant uncertainties in the parton momentum distributions within the 
proton, ambiguities in QCD calculations, and systematic uncertainties in jet 
energy measurement, make it difficult to discover a signal. 
This is apparent from the recent CDF measurement of the inclusive jet cross
section~\cite{cdf-jet}, and the phenomenological papers which 
followed~\cite{ref_partons}. Some progress has been made at the Snowmass 
workshop on quantifying the uncertainties in the parton 
distributions~\cite{ref_olness}, however, more work is clearly necessary.
Another signal of a $qq \rightarrow qq$ contact interaction, which is not
very sensitive to theoretical or jet energy measurement, is a dijet angular 
distribution which is more isotropic than predicted
by QCD.  Using 110 pb$^{-1}$ of data at the Tevatron, CDF has recently measured 
the dijet angular distribution and found good agreement with QCD predictions, 
thereby excluding 
a contact interaction among up and down type quarks with scale 
$\Lambda^+ \leq 1.6$ TeV or $\Lambda^- \leq 1.4$ TeV at 95\% 
CL~\cite{ref_dijet_angle}. For 
a flavor symmetric contact interaction among all quarks the exclusions are 
$0.2$ TeV more. Although with further luminosity this exclusion will 
improve somewhat, comparing this limit with that obtained from the UA1
experiment~\cite{ref_UA1}, we see that the compositeness scale reach of a 
hadron collider is roughly equal to its center of mass energy, $\sqrt{s}$. 
This is confirmed by studies of the LHC~\cite{hinchliffe}, where the predicted 
reach is $\Lambda \approx 15$ TeV. 

\subsubsection{$q\bar{q} \rightarrow \gamma\gamma$ Contact Interaction}

Rizzo has previously studied the effects of a 
$q\bar{q}\rightarrow \gamma \gamma$ contact interaction at the Tevatron
and the LHC~\cite{rizzo-prd}  and here he extends these results to a Very
Larger Hadron Collider (VLHC)~\cite{qqgamma}.
The lowest dimension gauge invariant operator involving two 
fermions and 
two photons is a dimension-8 operator, which induces a $q\bar 
q\gamma\gamma$ contact
interaction.   This interaction, assuming parity and CP conservation, is 
given by
\begin{equation}
\label{gamma}
{\cal L} = \frac{2ie^2}{\Lambda^4} Q_q^2 F^{\mu\sigma} F_\sigma^\nu \;
 \bar q \gamma_\mu \partial_\nu q \;,
\end{equation}
where $e$ is the electromagnetic coupling, and $\Lambda$ is the associated mass
scale.
\begin{figure}[tbh]
\vspace*{1in}
\includegraphics{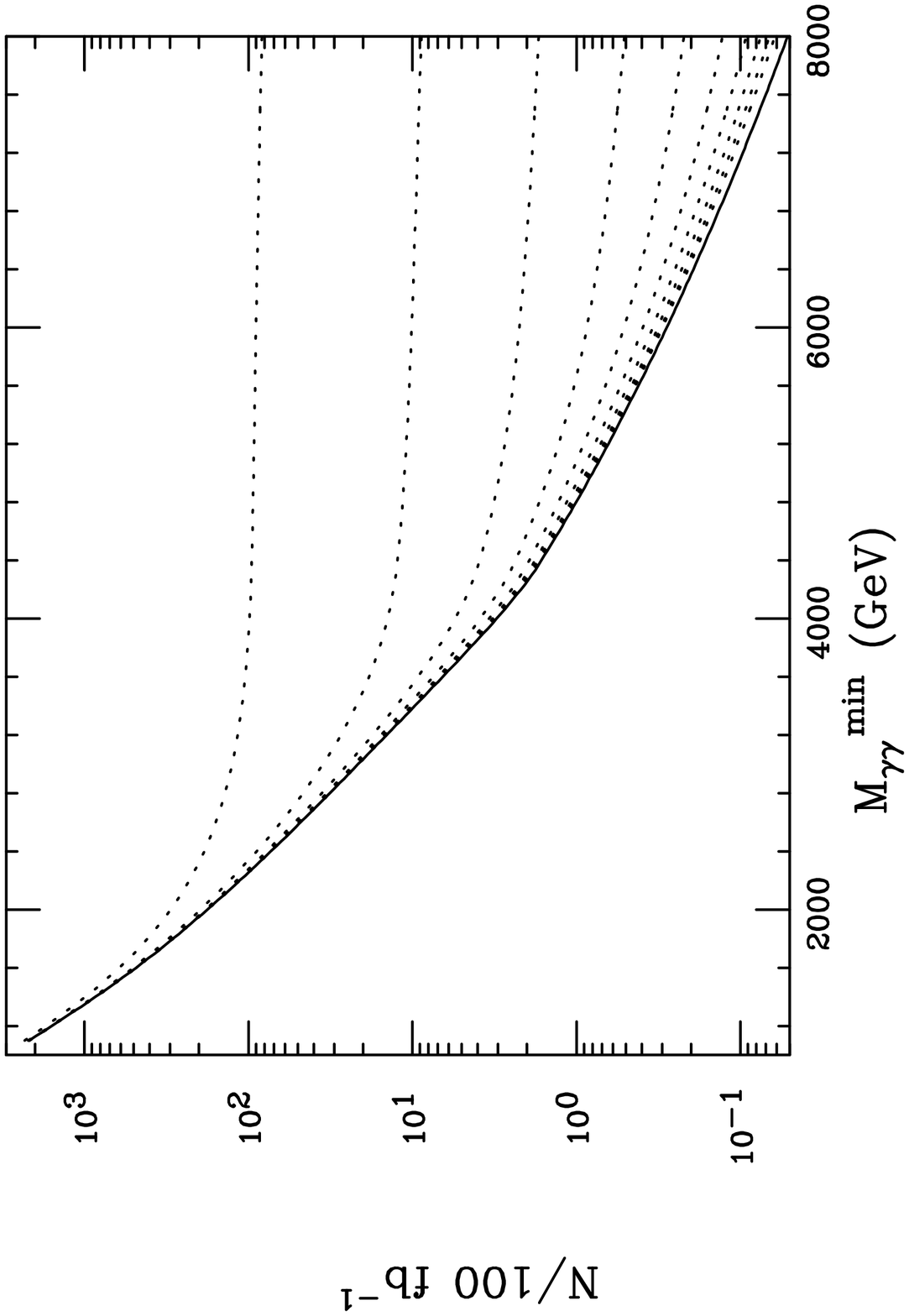}
\includegraphics{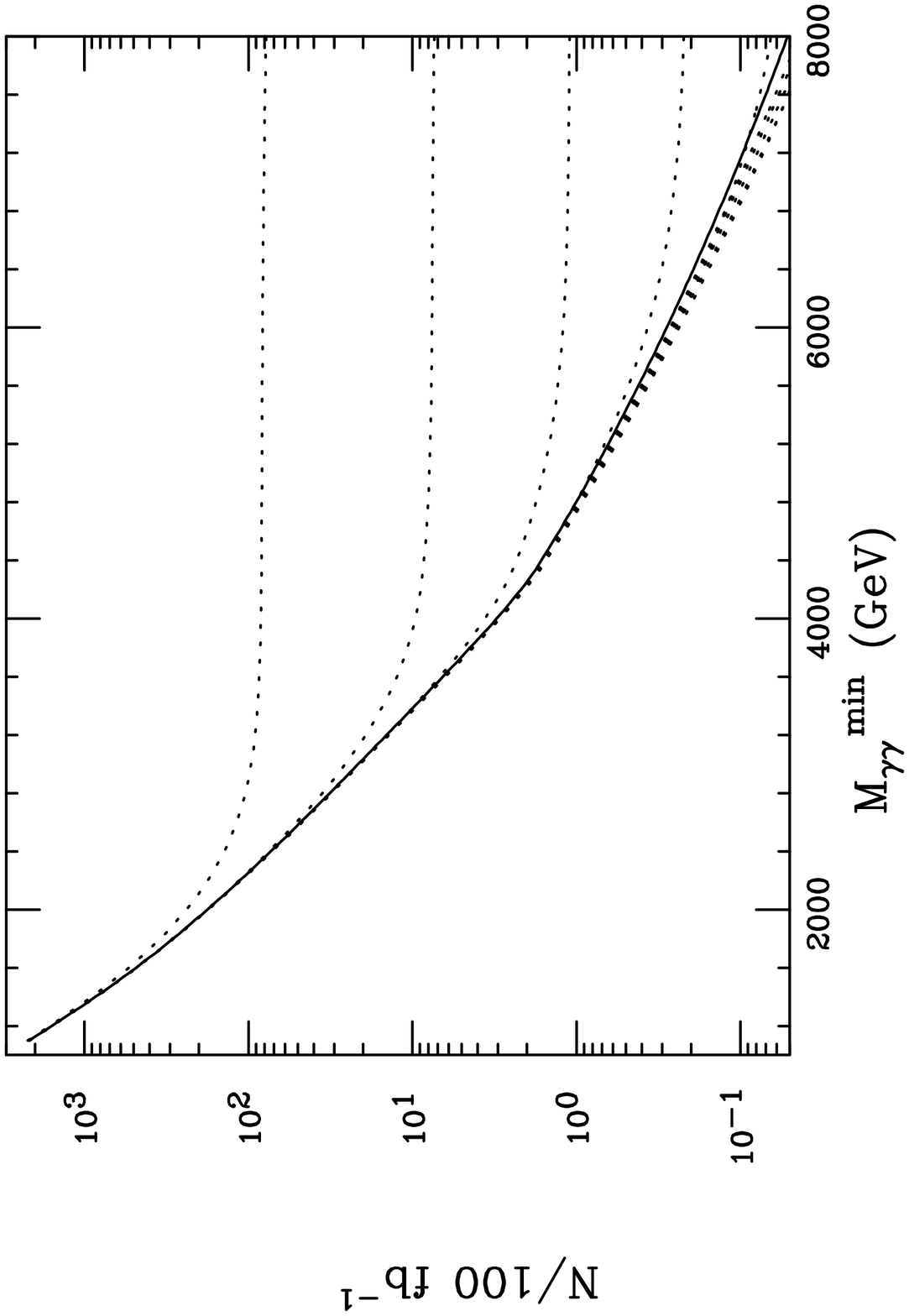}
\caption{Event rate for isolated $\gamma\gamma$ events with invariant masses
larger than $M_{\gamma\gamma}^{min}$at a 60 TeV $pp$ collider scaled to a 
luminosity of $100\;{\rm fb}^{-1}$.  The solid curves is the SM case while the 
top dotted curve corresponds to $\Lambda_+ (\Lambda_-)=3$ TeV in the left 
(right) figure.  Each subsequent dotted curve corresponds to an increase in
$\Lambda_\pm$ by 1 TeV.  In either case we have applied the cuts 
$p_t^\gamma \ge 500$ GeV and $|\eta_\gamma|\le 1$.}
\label{gamma-60}
\end{figure}
The observation of the signatures associated with this operator would be a clear
signal of compositeness.  
%
The mass scale 
$\Lambda^\pm$ indicates that the limits obtained below will depend upon
whether the contact operator interferes constructively or destructively
with the SM contribution.  
It is clear that the contact interaction in (\ref{gamma}) affects the 
parton cross section most in the region with large $\hat s$, and thus it 
causes the cross section to be less peaked in the forward and backward directions
and generates more central and higher $p_T$ photons.  It also
enhances the production rate at high diphoton invariant mass 
$M_{\gamma\gamma}$.

Figure \ref{gamma-60} shows the integrated event rates for isolated diphoton
events with invariant mass larger than $M_{\gamma\gamma}^{min}$ at a 
60 TeV $pp$ collider with a $100\; {\rm fb}^{-1}$ luminosity.  It clearly 
shows that the contact interaction of Eq.(\ref{gamma}) changes the cross
sections most in the high $M_{\gamma\gamma}^{min}$ region. 
In order to obtain the sensitivity to the contact interaction, we can 
assume that there is no event excess over the SM predictions in various
future collider experiments, and then we can put limits on $\Lambda^\pm$
using a simple $\chi^2$ analysis.  The results for various future 
collider experiments are tabulated in Table~\ref{tab_gamma}. 
From the table we can see that $p\bar p$ colliders are better than $pp$
colliders because there are more $q\bar q$ luminosities in $p\bar p$ than
in $pp$.  The limits can be pushed to about 7--13 TeV at a 60 TeV machine,
and about 16--33 TeV at a 200 TeV one.

\begin{table}[bth]
\begin{center}
\caption{$95\%$ CL bounds on the scale of the $q\bar q \gamma\gamma$ 
contact interaction at future hadron colliders. 
Here, $p_t^{min}$ is the minimum transverse momentum of each of the photons in 
GeV, ${\cal L}$ is the machine integrated 
luminosity in $fb^{-1}$, and 
$\Lambda^{\pm}$ is the lower bound on the scale in TeV.}
\label{tab_gamma}
\begin{tabular}{lccccc}
\hline
\hline
Machine& $p_t^{min}$& $|\eta_{\gamma,max}|$& $\cal L$& $\Lambda^+$& 
$\Lambda^-$ \\
\hline
TeV                 &  15 & 1 & 2 & 0.75 & 0.71 \\
LHC                 & 200 & 1,2.5 & 100 & 2.8 & 2.9 \\ 
60 TeV ($pp$)       & 500 & 1 & 100 & $\simeq 9.5$ & $\simeq 6.5$ \\
60 TeV ($p\bar p$)  & 500 & 1 & 100 & $\simeq 13.5 $ & $\simeq $ 10.5\\
200 TeV ($pp$)      &1000 & 1 & 1000 & $\simeq 23$ & $\simeq 16$ \\
200 TeV ($p\bar p$) &1000 & 1 & 1000 & $\simeq 33$ & $\simeq 26 $ \\ 
\hline
\hline
\end{tabular}
\end{center}
\end{table}

\vspace*{-.3in}
\subsection{Excited Quarks}

\begin{figure}[tbh]
\vspace*{-0.25in}
\hspace*{+0.1in}
\epsfysize=3.0in
\epsffile{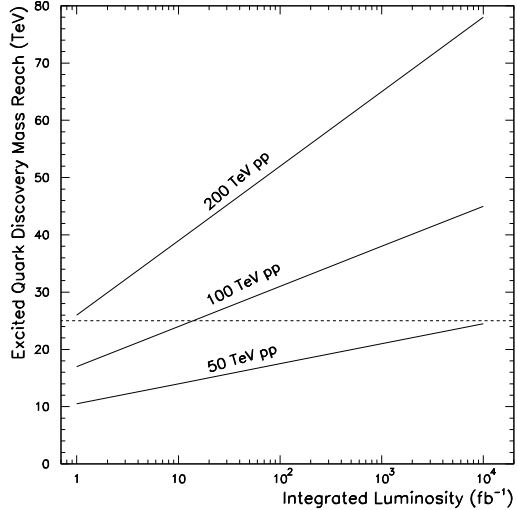}
\caption[]{ 
The $5\sigma$ discovery mass reach, for excited quarks decaying to dijets, is 
shown as a function of integrated luminosity for a VLHC with $\sqrt{s}=50$ TeV, 100 
TeV and 200 TeV (solid curves).  The horizontal dashed line demonstrates what 
luminosity is necessary to discover a 25 TeV excited quark.}
\label{fig_vlhc}
\end{figure}

Although it is expected that the first evidence for quark and/or lepton 
substructure would arise from the affects of contact interactions, 
conclusive evidence would be provided by observation of excitations 
of the preon bound state. If quarks are composite particles then excited 
quarks are expected.
Harris~\cite{ref_snow_exq} has investigated the prospects for discovering an
excited quark~\cite{ref_qstar}, $u^*$ 
or $d^*$ with spin 1/2 and weak isospin 1/2, at hadron colliders. 
He considers the process $qg\rightarrow q^* \rightarrow qg$, and 
does a lowest order calculation of the dijet resonance signal and QCD 
background assuming an experimental dijet mass resolution of 10\%. 
The estimated
$5\sigma$ discovery mass reach at the Tevatron is $0.94$ TeV for Run II
(2 fb$^{-1}$) and $1.1$ TeV for TeV33 (30 fb$^{-1}$).  The mass reach at the
LHC is $6.3$ TeV for 100 fb$^{-1}$.  The discovery mass reach at a Very
Large Hadron Collider (VLHC) is shown in Fig.~\ref{fig_vlhc} for 3 different
machine energies as a function of integrated luminosity. At a VLHC with a 
center of mass energy of (50) 200 TeV the mass reach is 25 TeV (78 TeV) 
for an integrated luminosity of $10^4$ fb$^{-1}$.  However, an excited quark 
with a mass of 25 TeV would be discovered at a hadron collider with 
$\sqrt{s}=100$ TeV and an integrated luminosity of only 13 fb$^{-1}$: here a 
factor of 2 
increase in energy from a 50 TeV to a 100 TeV machine is worth a factor of 
$1000$ increase in luminosity at a fixed machine energy of 50 TeV. 

\vspace*{-.3in}
\section{Anomalous Couplings of Quarks}

The lowest order interaction between a quark and a gluon is a dimension-4
operator, $\bar t \gamma^\mu T_a t G^a_\mu$. 
Among all the dimension-5 operators, the most interesting ones
involving quarks and gluons 
are the chromomagnetic (CMDM) and chromoelectric (CEDM) 
dipole moment couplings of quarks.  
These dipole moment couplings are important not only because they are 
only suppressed by one power of $\Lambda$ but also because a nonzero 
value for the CEDM is a clean signal for CP violation.  
The effects of these anomalous couplings have been studied quite extensively,
e.g., in $t\bar t$ production \cite{rizzo,haberl,cheung}, in $b\bar b$
production \cite{rizzo}, and in inclusive jet production \cite{dennis}.

The effective Lagrangian for the interactions between a quark and a gluon
that include the CEDM and CMDM form factors is 
\begin{equation}
\label{eff}
{\cal L}_{\rm eff} = g_s \bar q T^a \left[ -\gamma^\mu G_\mu^a +
\frac{\kappa}{4m_q} \sigma^{\mu\nu} G_{\mu\nu}^a - 
\frac{i \tilde{\kappa}}{4m_q} \sigma^{\mu\nu} \gamma^5 G_{\mu\nu}^a 
\right ] q \;.
\end{equation}
where $\kappa/2m_q$ ($\tilde{\kappa}/2m_q$) is the CMDM (CEDM) of the quark
$q$.  The above Lagrangian is valid for both light and heavy quarks. 
The Lagrangian in Eq.~(\ref{eff}) gives an effective $qqg$ vertex, and also
induces a $qqgg$ interaction, which is absent in the SM.
We shall use a short-hand notation:
\begin{equation}
\kappa' = \frac{\kappa}{2m_q}\;, \tilde{\kappa}' = \frac{\tilde{\kappa}}{2m_q}
\end{equation}
which are given in units of (GeV)$^{-1}$.

\subsection{Prompt Photon Production}

\begin{figure}[ht]
\vspace*{1.4in}
\includegraphics{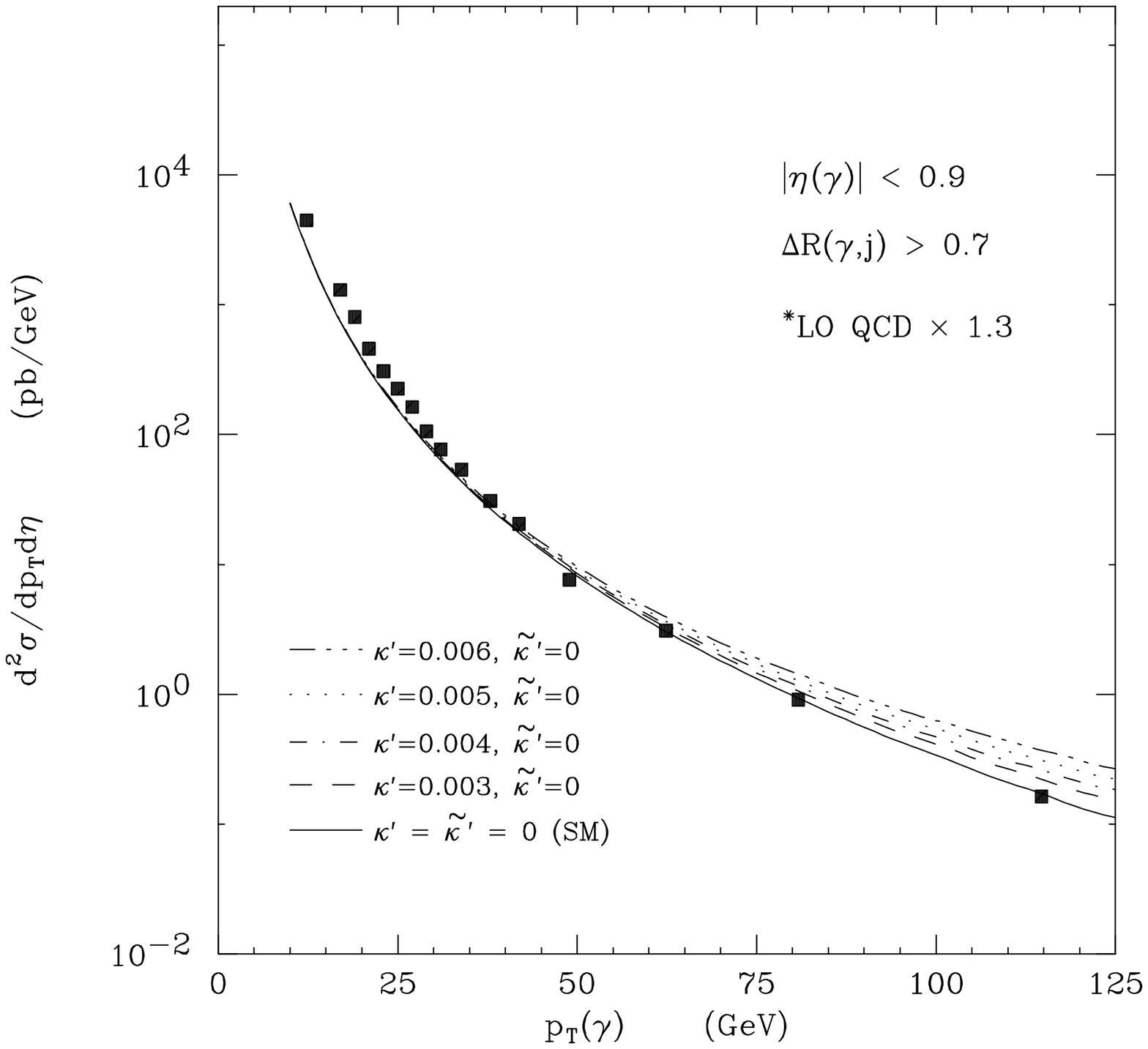}
\includegraphics{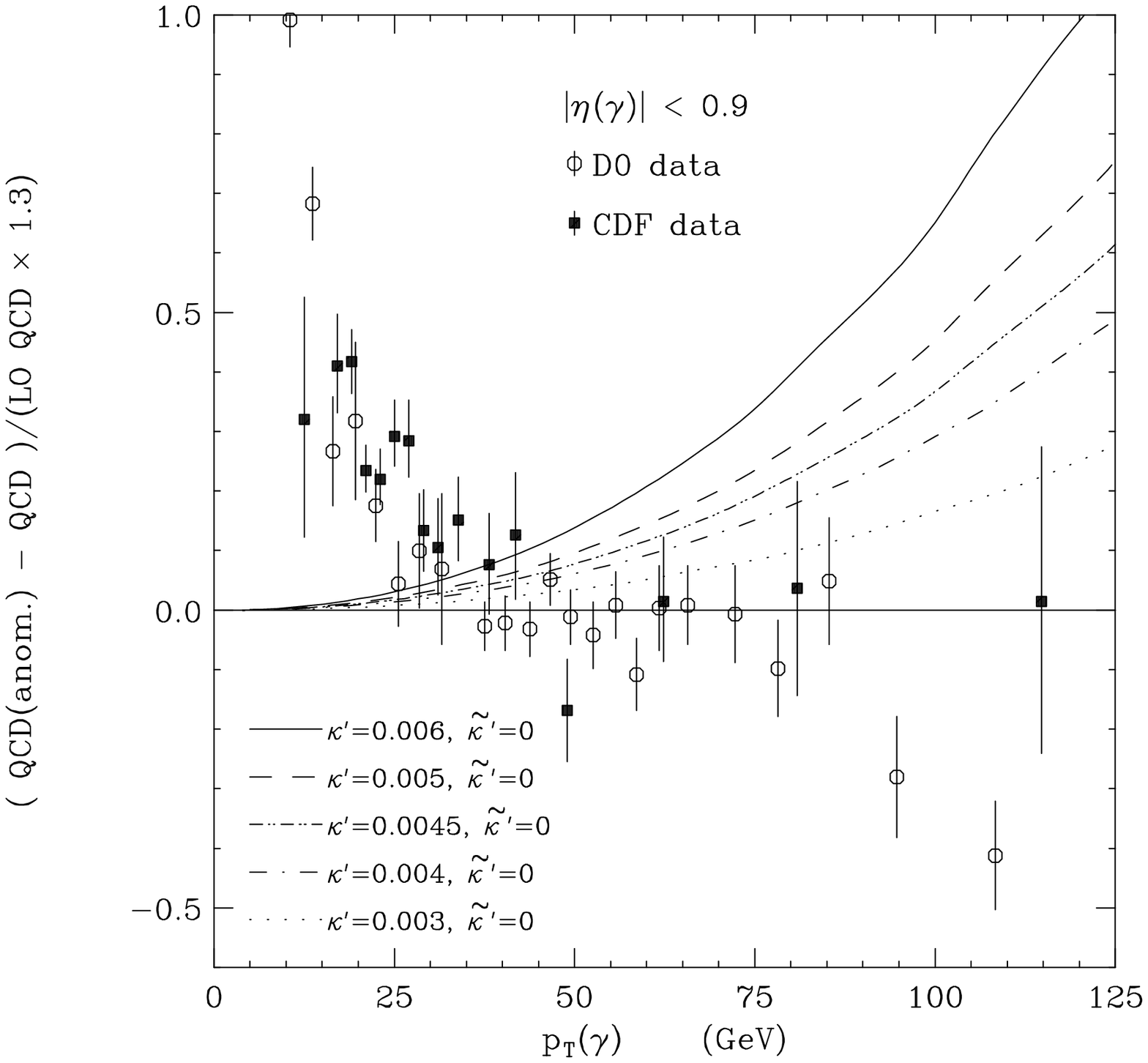}
\caption{
(a) Transverse momentum distribution $d^2\sigma/p_T d\eta$ in prompt photon 
production for pure QCD and nonzero values of $\kappa'$. The data points 
are from CDF;
(b) fractional difference from QCD for various values of $\kappa'$ and
$\tilde{\kappa}'$.  Both D0 and CDF data are shown.}
\label{diff}
\end{figure}

Cheung and Silverman have studied the effects
of anomalous CMDM and CEDM of light quarks on prompt photon production
\cite{photon}.
Prompt photon production is sensitive 
to the gluon luminosity inside a hadron 
because it is mainly produced by quark-gluon scattering.
For the same reason this process is also sensitive to the anomalous 
couplings of quarks to gluons.
The contributing subprocesses for prompt photon production are:
$q(\bar q) g \to \gamma q(\bar q)$ and $q\bar q \to \gamma g$.
The spin- and color-averaged amplitude for $q(p_1) g(p_2) \to \gamma(k_1) 
q(k_2)$ is given by
\begin{equation}
\label{amp1}
\overline{\sum} |{\cal M}|^2 = \frac{16\pi^2 \alpha_s
\alpha_{\rm em} Q_q^2}{3} \biggr [ - \frac{s^2 + t^2}{st} - 2 u 
(\kappa^{'2} + \tilde{\kappa}^{'2} ) \biggr]
\end{equation}
where
\begin{equation}
s=(p_1+p_2)^2\,, \;\; t=(p_1-k_1)^2 \,, \;\; u=(p_1-k_2)^2 \,,
\end{equation}
and $Q_q$ is the electric charge of the quark $q$ in units of proton charge.
Similarly, the spin- and color-averaged amplitude for 
$q(p_1) \bar q (p_2) \to \gamma(k_1) g(k_2)$ is given by
\begin{equation}
\label{amp2}
\overline{\sum} |{\cal M}|^2 = \frac{128\pi^2 \alpha_s
\alpha_{\rm em} Q_q^2}{9} \biggr [ \frac{t^2 + u^2}{ut} + 2 s 
(\kappa^{'2} + \tilde{\kappa}^{'2} ) \biggr] \;.
\end{equation}
The differential cross section for prompt photon production versus 
the transverse momentum of the photon is shown in Fig.~\ref{diff}a. 
The LO QCD curve has to be multiplied by a $K$-factor of about 1.3 to 
best fit the CDF data.  Figure~\ref{diff}a also shows curves with nonzero 
values of CMDM.  It is clear  that nonzero $\kappa'$ 
will increase the total and the differential cross sections, especially in
the large $p_T(\gamma)$ region.
The effects due to nonzero CEDM will be the same because the increase in
cross section is proportional to $(\kappa^{'2} + \tilde{\kappa}^{'2})$.

The fractional difference from pure QCD for nonzero CMDM is shown 
in Fig.~\ref{diff}b.  The data are from CDF \cite{cdf} and D0 \cite{d0}.  
The anomalous behavior at low $p_T(\gamma)$ has already been resolved 
by including initial and final state shower radiation, therefore, 
only the large $p_T$ region is relevant.
Since in Eqs.~(\ref{amp1}) and (\ref{amp2}) the role of 
$\kappa'$ and $\tilde{\kappa}'$ are the same,  one of them is kept zero
when bounding on the other.   From these curves it is clear that the CDF 
and D0 data would be inconsistent with $\kappa' > 0.0045$, therefore, 
giving a bound of 
\begin{equation}
\label{bound}
\kappa' \le 0.0045 \;\;{\rm GeV}^{-1}
\end{equation}
on the CMDM of light quarks.  
Similarly, a bound of $\tilde{\kappa}' \le 0.0045 \;\;{\rm GeV}^{-1}$
on the CEDM of light quarks is valid as well.  
We compare this with the results obtained in 
Ref.\cite{dennis} for jet production.  The value of $\kappa'$ obtained in fitting to the 
CDF\cite{cdf-jet} transverse energy distribution of the inclusive jet 
production is\cite{dennis}
$\kappa' = (1.0 \pm 0.3 )\times 10^{-3} \;\; {\rm GeV}^{-1}$
which is consistent with the bound in Eq.~(\ref{bound})

\vspace*{-.1in}
\subsection{Sensitivities in Future Collider Experiments}

Silverman and Cheung~\cite{project} have estimated the sensitivity of the
Tevatron and LHC to 
the anomalous chromomagnetic dipole moment of light quarks.
A lowest order parton level calculation was used, and only the statistical
sensitivity of the experiments was considered.
The criterion, in the spirit of reference~\cite{hinchliffe},
is to take bins of appropriate size for the energy range being
examined, and find the $E_T$ called $E_T^*$ at which the QCD cross
section statistical error bars are 10\%.  These will be bins with
100 QCD events.  Then the cross section due to QCD plus the anomalous 
chromomagnetic moment contribution will be explored, and 
the value of $\kappa' \equiv 1/\Lambda$ or $\Lambda$ is determined 
where the excess over QCD is 10\% at this $E_T^*$.  
These $E_T^*$ and $\Lambda$ are shown in Table~\ref{tab_coupling}.  
Varying
the bin size by a factor of two makes only a small change in the value
of $E_T^*$ or $\Lambda$.  The limits in $|\eta|$ used are 0.9
for the Tevatron, and 1.0 for the LHC.
From table~\ref{tab_coupling} one can see that $\Lambda$ sensitivity scales roughly  
as the beam energy.

\begin{table}[th]
\vspace*{-.2in}
\begin{center}
\caption{Table of High $E_T$ Bins at 10\% Statistical Error and
1-$\sigma$ Sensitivity for $\Lambda$ in that Bin, is shown as a
function of machine energy, integrated luminosity, and bin width.}
\label{tab_coupling}
\begin{tabular}{|r|r|r||r|r||r|r|}
\hline
 & Int. & Bin &\multicolumn{2}{c|| }{$E_T$ Jets}  
&\multicolumn{2}{c|}{Photons} \\
\cline{4-7} \cline{4-7}
$E_{\rm cm}$ & Lum. & Width
& $E_T^*$ & $\Lambda$ & $E_T^*$ & $\Lambda$ \\
\hline
 TeV & fb$^{-1}$ & GeV & GeV & TeV & GeV & TeV \\
\hline
         1.8 & 0.1 & 10 & 360 & 1.8 & 140 & 0.7 \\
         2.0 & 2   & 20 & 490 & 2.8 & 260 & 1.5 \\
         2.0 & 10  & 20 & 540 & 3.3 & 325 & 1.9 \\
         2.0 & 30  & 20 & 575 & 3.5 & 370 & 2.1 \\
\hline
         14  & 10  & 100 & 2500 & 13 & 1000 & 4.5 \\
         14  & 100 & 100 & 3100 & 17 & 1400 & 6.3 \\
\hline
\end{tabular}
\end{center}
\end{table}

\subsection{Effects on $t\bar t$ Production at the LHC}

Top quark production at hadronic colliders is the most obvious place to
probe the anomalous coupling of top quarks to gluons.  There have been quite a 
few studies \cite{rizzo,haberl,cheung} on this subject at the Tevatron energies.
Rizzo has extended the study to the LHC 
\cite{lhc}.  The contributing subprocesses to top pair production
are $q\bar q,\; gg \to t\bar t$. The existence of a nonzero chromomagnetic
dipole moment of the top quark will change both the total and differential 
cross sections.  Since higher partonic 
center-of-mass energies become accessible at the LHC , one can probe beyond 
the top pair production threshold region, and have much higher sensitivities 
to the CMDM.

\begin{figure}[hbt]
\vspace*{-.1in}
\includegraphics{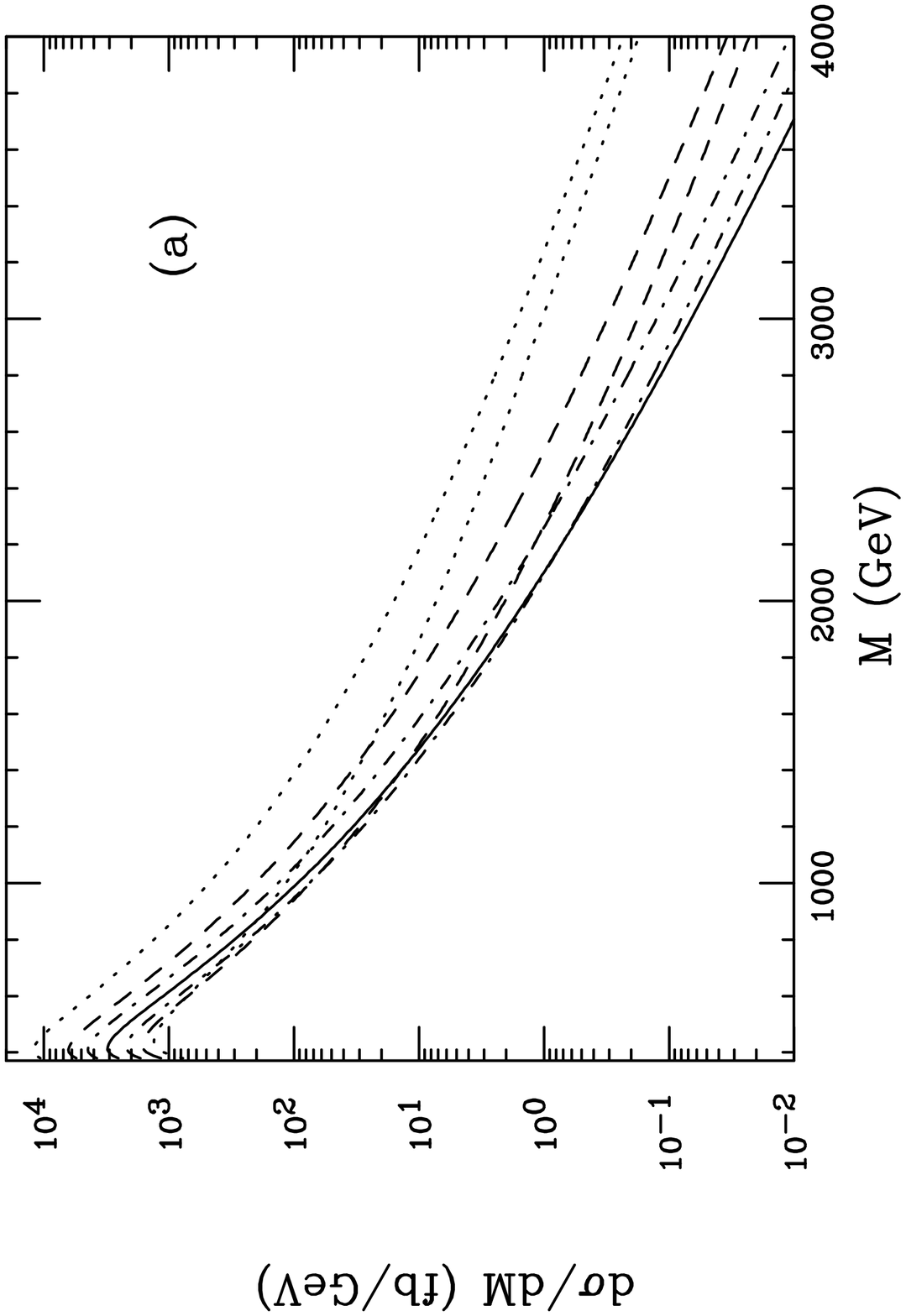}
\includegraphics{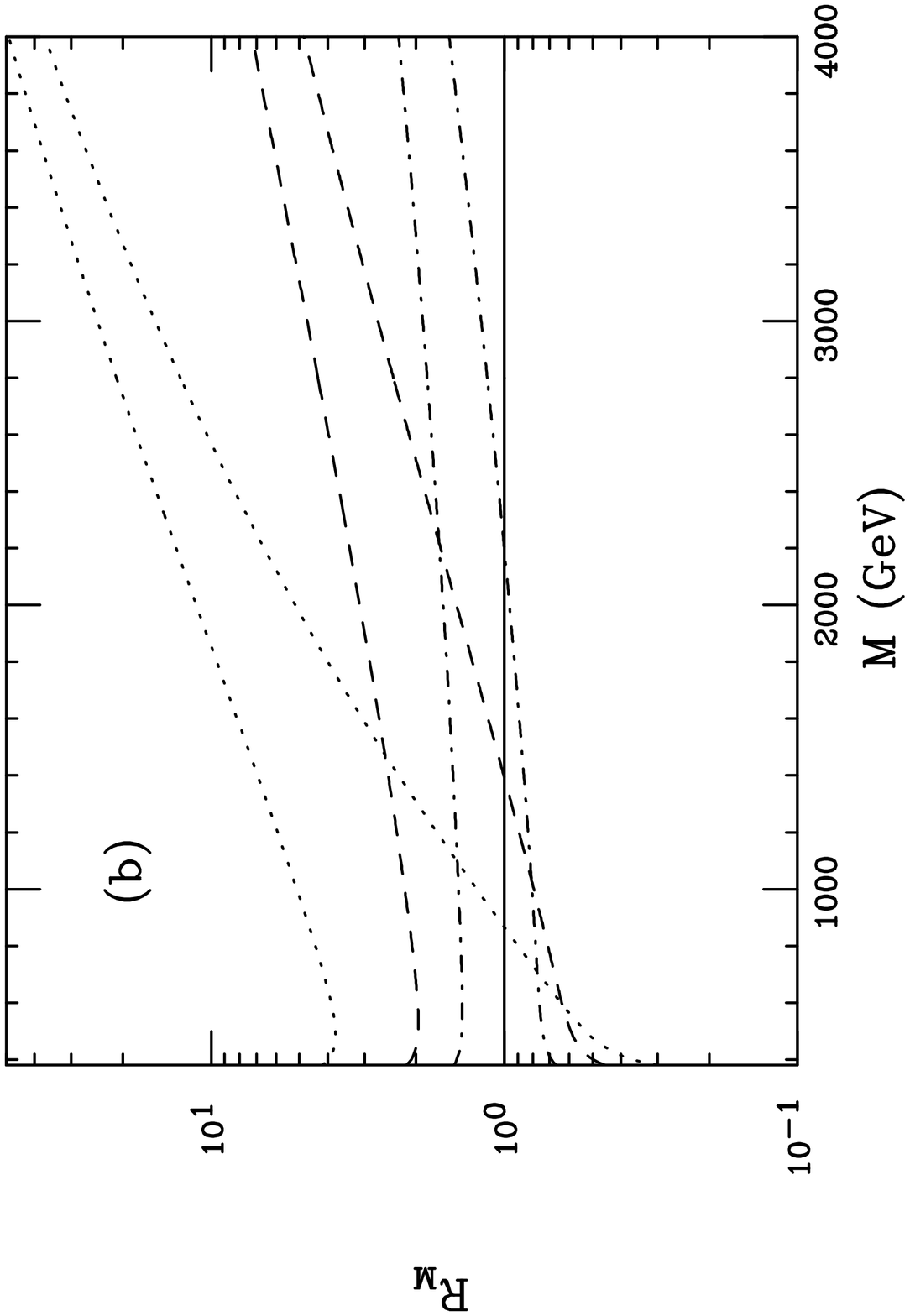}
\includegraphics{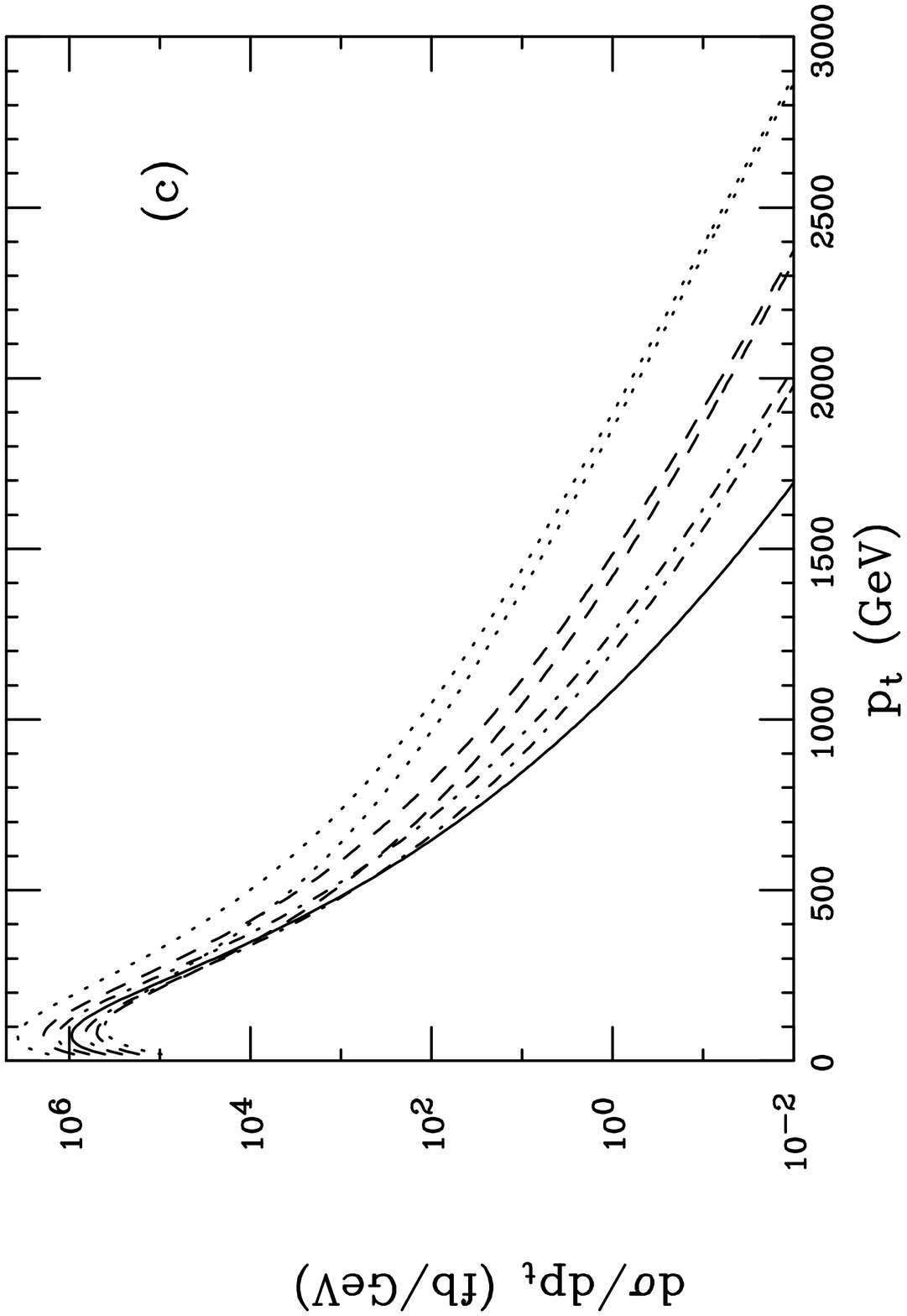}
\includegraphics{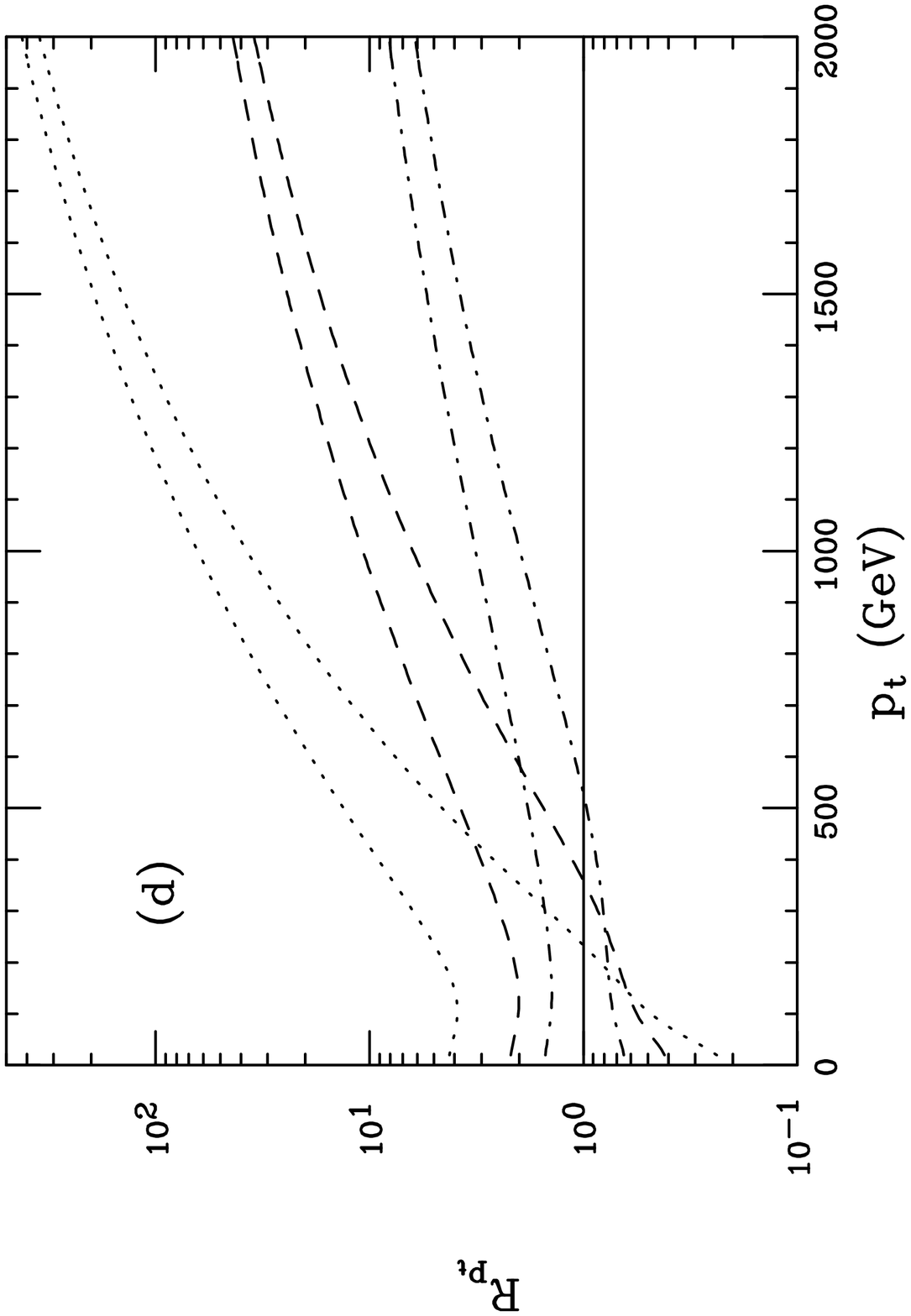}
\vspace*{2.8in}
\caption{(a) $t\bar t$ invariant mass distribution at the LHC for 
various values of $\kappa$ assuming $m_t=180$ GeV. (b) The same distribution 
scaled to the SM result. (c) $t\bar t$ $p_t$ distribution at the LHC and (d) 
the same distribution scaled to the SM. In all cases, the SM is represented 
by the solid curve whereas the upper(lower) pairs of dotted, dashed, and
dash-dotted curves corresponds to $\kappa=$0.5(-0.5), 0.25(-0.25), and 
0.125(~-0.125),  respectively.}
\label{dist}
\end{figure}

Figures~\ref{dist}a and \ref{dist}c show the 
modifications in the SM expectations for both $d\sigma/dM_{tt}$ and 
$d\sigma/dp_t$, respectively, for different values of $\kappa$ of the top 
quark. 
Perhaps more revealing, Figures~\ref{dist}b and~\ref{dist}d show the 
ratio of the modified distributions to the corresponding SM ones.  One 
can see that a non-zero $\kappa$ leads to ($i$) enhanced cross 
sections at large $p_t$ and $M_{tt}$, and ($ii$) the {\it shapes} of the 
distributions are altered, {\it i.e.}, the effect is not just an overall 
change in normalization.
The sensitivities of these distributions to nonzero $\kappa$ are also estimated
using a Monte Carlo approach, taking into account a reasonable size of 
systematic errors.  Assuming the SM is the correct theory, the 
95\% CL allowed regions of $\kappa$ of the top quark
are $-0.09 \le \kappa \le 0.10$ from
the $M_{tt}$ distribution and $0.06 \le \kappa \le 0.06$ from the 
$p_T$ distribution. 

\onecolumn

\begin{table}[th]
\begin{center}
\caption{The mass reach and discovery potential, for particles from one-family
technicolor or topcolor, at the Tevatron as a function of integrated 
luminosity}
\label{tab_tev}
\begin{tabular}{|c|c|c|c|} \hline
Channel & Run I & Run II & TeV33 \\ 
        & (.1 fb$^{-1}$) & (2 fb$^{-1}$) & (30 fb$^{-1}$) \\ \hline
$\rho_{T1} \rightarrow W$ + dijet & No Mass & No Mass & $\approx 400$ GeV \\
(no b-tagging)                    & Reach   &  Reach  & at 95\% CL \\ \hline
$\rho_{T1} \rightarrow W$ + dijet &   & Discovery for & Better than  \\
(with b-tagging)                  &  ?  & $M_\rho=210$ $M_\pi=115$ GeV & no b-tag\\ \hline
$\rho_{T8} \rightarrow$ dijet & $0.25 < M_\rho < 0.50$ TeV & $0.77$ TeV & $0.90$ TeV \\
                 & at 95\% CL               &  at 95\% CL & at 95\% CL \\ \hline
Topgluon $B^{\prime}\rightarrow b\bar{b}$ & Search in & $0.77-0.95$ TeV & $1.0-1.2$ TeV \\
($0.3<\Gamma/M<0.7$)                     & Progress & at $5\sigma$ & at $5\sigma$\\ \hline
Topgluon $B^{\prime}\rightarrow t\bar{t}$ & Search in & $0.97-1.11$ TeV & $1.3-1.4$ TeV \\
($0.3<\Gamma/M<0.7$)                     & Progress & at $5\sigma$ & at $5\sigma$\\ \hline
TopC $Z^{\prime}\rightarrow t\bar{t}$ & Search in & 920 GeV &  1150 GeV \\
($\Gamma/M=.012$)                     & Progress & at $5\sigma$ & at $5\sigma$\\ \hline
\end{tabular}
\end{center}
\end{table}

\begin{table}[bh!]
\vspace*{-.2in}
\Large
\begin{center}
\caption{Mass and energy reach in TeV for new interactions at colliders. 
The symbol 
``$\sim$'' indicates a guess based on scaling from lower energy machines. 
``Found'' indicates the collider will discover the particle if it exists, and 
``Already Found'' indicates the particle would have already been discovered 
by a earlier collider. The symbol ``--'' means not applicable, and ``?'' means 
we don't know. The numbers in square brackets are either confidence levels in 
\% or RMS deviations in units of $\sigma$, indicating the statistical size of 
the effect corresponding to the mass or energy reach.}
\label{tab_summary}
\hspace*{-.1in}
\begin{tabular}{|c||c|c|c||c|c|} \hline
 Particle & \multicolumn{5}{c|}{Collider} \\ \cline{2-6}
or & TeV33 & LHC & VLHC & NLC & Muon \\
Interaction  & 2 TeV, $p\bar{p}$ & 14 TeV, $pp$ & 200 TeV, $pp$  & .5 TeV, $e^+e^-$ & 
4 TeV, $\mu^+\mu^-$ \\ 
Scale & 30 $fb^{-1}$ & 100 $fb^{-1}$ & 1000 $fb^{-1}$ & 50 $fb^{-1}$ & 1000 $fb^{-1}$  \\ \hline \hline
Technicolor $\rho_{T1}$ & 
.4 {\normalsize [95\%]} & $>1^{\ast}$ & ? &  
$1.5^{\dagger}$ {\normalsize [$6.7\sigma$]} & $\sim 10$ \\ \hline
Techni $\rho_{T8}\rightarrow$ dijet& Found & \multicolumn{4}{c|}{Already Found} \\ \hline
Topcolor $Z^{\prime}\rightarrow t\bar{t}$ & 1.1 {\normalsize [$5\sigma$]}
 & Found & \multicolumn{3}{c|}{Already Found} \\ \hline
Topgluon $B\rightarrow b\bar{b},t\bar{t}$ & 1.4 {\normalsize [$5\sigma$]} 
& Found & \multicolumn{3}{c|}{Already Found} \\ \hline
$\Lambda_{LL}(l\bar{l}\rightarrow l^{\prime}\bar{l}^{\prime})$ & -- & -- & -- & 
$19$ {\normalsize [95\%]} & 110 {\normalsize [95\%]} \\ \hline
$\Lambda_{LL}(qq\rightarrow qq)$ & $2$ & $15^{\ast}$ & $\sim 200$ & -- & -- \\ \hline
$\Lambda_{LL}(q\bar{q}\leftrightarrow l\bar{l})$ & 20 {\normalsize [95\%]}& $\sim 100$ & $\sim 1000$ & 
24 {\normalsize [95\%]} & 140 {\normalsize [95\%]}\\ \hline
$\Lambda(q\bar{q}\rightarrow \gamma \gamma)$ & 0.9 
{\normalsize [95\%]}& 3 {\normalsize [95\%]} & 20 {\normalsize [95\%]} & -- & -- \\ \hline
Excited Quark & 1.1 {\normalsize [$5\sigma$]} & 
6.3 {\normalsize [$5\sigma$]} & 78 {\normalsize [$5\sigma$]} 
& $0.45^{\dagger}$ & $\sim 3$ \\ \hline
CMDM $\Lambda$ (dijets) & 3.5 {\normalsize [$>1\sigma$]}  & 
17 {\normalsize [$>1\sigma$]} & ? & -- & -- \\ \hline
CMDM $\Lambda$ ($\gamma$+jet) & 2.1 {\normalsize [$>1\sigma$]} & 
6.3 {\normalsize [$>1\sigma$]} & ? & -- & -- \\ \hline
\end{tabular} \\
$^\ast$ from reference \cite{hinchliffe} \ \  
$^\dagger$ from reference \cite{ref_nlc_zdr} 
\end{center}
\end{table}
\twocolumn

\subsection{Summary and Conclusions}

Table~\ref{tab_tev} and ~\ref{tab_summary} summarize the ability of colliders
to answer fundamental questions involving new interactions.

Is electroweak symmetry broken by the dynamics of a new interaction? 
We see that a color singlet technirho can be discovered at the 
Tevatron if it has the
mass expected within the one-family technicolor model.  Simpler models of 
technicolor, where there are only color singlet techniquarks and no 
technileptons, would predict technirho masses of a TeV or more.
These could be discovered at the LHC or NLC,
and higher mass technirhos could be observed
at a VLHC or a muon collider. Color octet technirhos can be discovered at 
the Tevatron if they have the mass expected within the one-family technicolor
model. 

Is the mass difference between the top quark and the other quarks 
generated by a new interaction? Topcolor assisted technicolor can be discovered
at the Tevatron if the topgluon or topcolor $Z^{\prime}$ has mass around a 
TeV or less, which is possible. The topgluon and topcolor $Z^{\prime}$ are
expected to be lighter than a few TeV, so if they are missed by the Tevatron 
they will be discovered by the LHC. 

Are quarks and leptons composite particles held together by new interactions?
If the energy scale of those interactions is less than 20 TeV, the Tevatron
has a chance of discovery in the dilepton mass spectrum, the NLC has
a slightly better chance of discovery using dijet angular distributions, and 
the LHC will certainly see this scale of $q\bar{q}\leftrightarrow l\bar{l}$ 
contact interaction.
Proof that observed contact interactions were caused by compositeness would
come from the observation of excited states with mass near the compositeness
scale. To discover an excited quark with mass around 20 TeV, we would have to 
build a VLHC colliding protons with $\sqrt{s}=50-200$ TeV. 

Is there a new interaction which changes the coupling of quarks and gluons
at high energies?  The Tevatron can probe anomalous coupling energy scales of
a few TeV, and the LHC can probe 17 TeV for light quarks and is sensitive to 
top quark anomalous couplings.

We conclude that there is a significant chance of discovering new interactions 
at the Tevatron in the next decade. From Table~\ref{tab_summary} the reader
can determine which of the proposed future colliders provide the greatest 
additional discovery potential in the post-Tevatron era.

%


\vspace*{-0.1in}
\begin{thebibliography}{99}
  
\bibitem{ref_lane} E. Eichten and K. Lane, Fermilab-Conf-96/297-T, BUHEP-96-33, 
hep-ph/9609297, and these proceedings.

\bibitem{ref_toback} D. Toback, Fermilab-Conf-96/360 and these proceedings.

\bibitem{ref_womersley} J. Womersley, these proceedings.

\bibitem{ref_dijet} F. Abe {\em et al.} (CDF Collaboration), Phys. Rev. Lett. {\bf 74}, 3538 (1995).

\bibitem{ref_tev2000} D. Amidei and R. Brock, {\em Report of the TeV2000 Study
Group}, Fermilab-Pub-96/082.

\bibitem{ref_taekoon} T. Lee, these proceedings.

\bibitem{ref_lane2} E. Eichten and K. Lane, Fermilab-Conf-96/298-T, 
BUHEP-96-34, hep-ph/9609298, and these proceedings.

\bibitem{ref_topcolor} C. Hill and S. Parke, Phys. Rev. {\bf D49}, 4454 
(1994).

\bibitem{ref_burdman} G. Burdman, these proceedings.

\bibitem{ref_hill} C. Hill, Phys. Lett. {\bf B345}, 483 (1995).

\bibitem{ref_topg_bbbar} R. Harris, Fermilab-Conf-96/276-E, hep-ph/9609316, and 
these proceedings.

\bibitem{ref_pbarp} R. Harris (CDF Collaboration), hep-ex/9506008, Fermilab-conf-95/152-E.

\bibitem{ref_topg_range} G. Buchalla, G. Burdman, C. Hill and D. Kominis,
Phys. Rev. {\bf D53}, 5185 (1996).

\bibitem{ref_topg_ttbar} R. Harris, Fermilab-Conf-96/277-E, hep-ph/9609318, and 
these proceedings.

\bibitem{ref_tollefson} K. Tollefson in reference~\cite{ref_tev2000}.

\bibitem{ref_pythia} T. Sjostrand, computer code PYTHIA V5.6, CERN-TH-6488/92.

\bibitem{ref_cdf_top} F. Abe {\em et al.} (CDF Collaboration), Phys. Rev. Lett. {\bf 74}, 2626 (1995).

\bibitem{ekp}E. Eichten, K. Lane, and M. Peskin, Phys. Rev. Lett. {\bf 50},
811 (1983).

\bibitem{cgh}K. Cheung, S. Godfrey, and J. Hewett, these proceedings.

\bibitem{ref_barbaro} P. de Barbaro {\em et al.}, Fermilab-Conf-96/356-E, and 
these proceedings.

\bibitem{cdf-jet}F. Abe {\em et al.} (CDF Collaboration), Phys. Rev. Lett. 
{\bf 77}, 438 (1996).

\bibitem{ref_partons} J. Huston {\em et al.}, Phys. Rev. Lett. {\bf 77}, 444 
(1996); H. L. Lai {\em et al.}, MSU-HEP-60426, hep-ph/9606399 (1996), submitted
to Phys. Rev. D.

\bibitem{ref_olness} F. Olness, these proceedings.

\bibitem{ref_dijet_angle} F. Abe {\em et al.} (CDF Collaboration), Fermilab-Pub-96/317-E, 
hep-ex/9609011, submitted to Phys. Rev. Lett.

\bibitem{ref_UA1} G. Arnison, {\em et al.}, Phys. Lett. B{\bf 177}, 244(1986)

\bibitem{hinchliffe} U.S. ATLAS and U.S. CMS Coll., edited by
I. Hinchliffe and J. Womersley, LBNL-38997 (1996).

\bibitem{rizzo-prd}T. Rizzo, Phys. Rev. {\bf D51}, 1064 (1994).

\bibitem{qqgamma}{\it Constraint on $qq\gamma\gamma$ Contact Interactions at
Future Hadron Colliders}, T. Rizzo, in these proceedings.

\bibitem{ref_snow_exq} R. Harris, Fermilab-Conf-96/285-E, hep-ph/9609319, and 
these proceedings.

\bibitem{ref_qstar} 
U.\ Baur, I.\ Hinchliffe and D.\ Zeppenfeld, Int. J. of Mod. Phys.
 {\bf A2}, 1285 (1987) and U.\ Baur, M.\ Spira and P.\ M.\ Zerwas, Phys. Rev.
 {\bf D42}, 815 (1990).

\bibitem{rizzo}D. Atwood, A. Kagan, and T. Rizzo, Phys. Rev. {\bf D52}, 6254
(1995).

\bibitem{haberl}P. Haberl, O. Nachtmann, and A. Wilch, Phys. Rev. {\bf D53},
4875 (1996).

\bibitem{cheung}K. Cheung, Phys. Rev. {\bf D53}, 3604 (1996).


\bibitem{dennis}D. Silverman, hep-ph/9605318, to be published in Phys. Rev.
{\bf D}.

\bibitem{photon}{\it Limits on Anomalous Couplings of Quarks by Prompt Photon
Data}, K. Cheung and D. Silverman, in these proceedings.

\bibitem{cdf}F. Abe {\em et al.} (CDF Collaboration), Phys. Rev. Lett. 
{\bf 73}, 2662 (1994).

\bibitem{d0}S. Abachi {\em et al.} (D0 Collaboration), FERMILAB-PUB-96/072-E.


\bibitem{project}{\it Quark Anomalous Chromomagnetic Moment Bounds - Projection
to Higher Luminosities and Energy}, K. Cheung and D. Silverman, in these
proceedings.

\bibitem{lhc}{\it Constraints on Anomalous Top Quark Couplings at the LHC},
T. Rizzo, in these proceedings.

\bibitem{ref_nlc_zdr} The NLC ZDR, Fermilab-Pub-96/112.

\end{thebibliography}
\end{document}